\documentclass[a4paper]{IEEEtran}
\usepackage[T1]{fontenc}
\usepackage[latin9]{inputenc}
\usepackage{color}
\usepackage{float}
\usepackage{amsmath}
\usepackage{amsthm}
\usepackage{amssymb}
\usepackage{stackrel}
\usepackage{graphicx}
\usepackage{esint}
\usepackage[unicode=true,
 bookmarks=false,
 breaklinks=false,pdfborder={0 0 1},backref=false,colorlinks=false]
 {hyperref}

\makeatletter

\pdfpageheight\paperheight
\pdfpagewidth\paperwidth

  \theoremstyle{plain}
  \newtheorem{thm}{\protect\theoremname}
  \theoremstyle{remark}
  \newtheorem{rem}{\protect\remarkname}

\usepackage{xcolor}
\usepackage{pdfcolmk}
\usepackage{units}

\usepackage{epsfig}
\usepackage{float}
\usepackage{subfigure}
\usepackage{subfloat}
\usepackage{cite}

\usepackage{miniplot}

\allowdisplaybreaks

\PassOptionsToPackage{normalem}{ulem}
\usepackage{ulem}

  \pdfpageheight\paperheight
\pdfpagewidth\paperwidth

\makeatother

\providecommand{\remarkname}{Remark}
\providecommand{\theoremname}{Theorem}

\begin{document}

\title{Capacity of Cooperative Vehicular Networks with Infrastructure Support:
Multi-user Case}

\author{Jieqiong Chen, \IEEEmembership{Student Member, IEEE}, Guoqiang Mao,
\IEEEmembership{Senior Member, IEEE}\\
Changle Li, \IEEEmembership{Senior Member, IEEE} , Weifa Liang, \IEEEmembership{Senior Member, IEEE}
and Degan Zhang }
\maketitle
\begin{abstract}
Capacity of vehicular networks with infrastructure support is both
an interesting and challenging problem as the capacity is determined
by the inter-play of multiple factors including vehicle-to-infrastructure
(V2I) communications, vehicle-to-vehicle (V2V) communications, density
and mobility of vehicles, and cooperation among vehicles and infrastructure.
In this paper, we consider a typical delay-tolerant application scenario
with a subset of vehicles, termed Vehicles of Interest (VoIs), having
download requests. Each VoI downloads a distinct large-size file from
the Internet and other vehicles without download requests assist the
delivery of the files to the VoIs. A cooperative communication strategy
is proposed that explores the combined use of V2I communications,
V2V communications, mobility of vehicles and cooperation among vehicles
and infrastructure to improve the capacity of vehicular networks.
An analytical framework is developed to model the data dissemination
process using this strategy, and a closed form expression of the achievable
capacity is obtained, which reveals the relationship between the capacity
and its major performance-impacting parameters such as inter-infrastructure
distance, radio ranges of infrastructure and vehicles, sensing range
of vehicles, transmission rates of V2I and V2V communications, vehicular
density and proportion of VoIs. Numerical result shows that the proposed
cooperative communication strategy significantly boosts the capacity
of vehicular networks, especially when the proportion of VoIs is low.
Our results provide guidance on the optimum deployment of vehicular
network infrastructure and the design of cooperative communication
strategy to maximize the capacity.
\end{abstract}


\section{Introduction}

Interest is surging on vehicular networks and connected vehicle technology
due to their increasingly important role in improving road traffic
efficiency, enhancing road safety and providing real-time information
to drivers and passengers \cite{Zheng15,Ilarri15,Li2014,Ning14}.
Two major wireless communication modes: vehicle-to-infrastructure
(V2I) communications and vehicle-to-vehicle (V2V) communications,
are supported in vehicular networks. V2V and V2I communications can
be realized by deploying wireless communication infrastructure points
along the roadside (e.g., road-side units (RSU), or LTE base stations),
equipping vehicles with on-board communication facilities (e.g., on-board
units (OBU)), and with the assistance of dedicated short-range communication
(DSRC) \cite{Kenny11,Morgan10} and LTE technology.

V2I and V2V communications, on one hand, are both major techniques
to disseminate data for vehicular applications, including safety applications
like disseminating real-time information about traffic accidents,
traffic congestion or obstacles in the road, and non-safety applications
such as offering value-added services (e.g., digital maps with real-time
traffic status) and in-car entertainment services \cite{Ilarri15,Li2014}.
On the other hand, as pointed out in the literature, purely relying
on V2I communications or V2V communications alone cannot meet the
diverse communication requirements of different vehicular applications.
For instance, V2V communications may become unreliable when the number
of hops in the communication becomes large \cite{Khabazian13}. They
may also not be supported and incur long communication delay when
the vehicular density is low \cite{Zhang2014,Du15,Shahidi}. Furthermore,
V2I communications may have limited availability, especially in rural
areas and in the initial deployment phase of vehicular networks due
to the high deployment cost. The aforementioned factors may result
in frequent interruptions in data transmissions, especially when downloading
files of large sizes from the Internet, e.g., in-car entertainment
services. Therefore, V2I and V2V communications have to co-exist and
complement each other to improve the network performance. 

Capacity is one of the most important performance metrics in vehicular
networks. Capacity of vehicular networks with infrastructure support
is both an interesting and challenging problem as the capacity is
determined by the inter-play of multiple factors including V2I communications,
V2V communications, density and mobility of vehicles, and cooperation
among vehicles and infrastructure. Since the seminal work of Gupta
and Kumar \cite{Gupta}, extensive research on capacity has been conducted,
e.g., \cite{Gross02,Mao13,Dai2016}. Focusing on the capacity of vehicular
networks, Wang et al. \cite{Wang14} studied urban vehicular networks
with uniformly distributed RSUs and analyzed the asymptotic uplink
throughput scaling law when the total number of vehicles is sufficiently
large. In \cite{Huang2015}, Huang et al. introduced a Euclidean planar
graph and used a practical geometric structure to study the asymptotic
capacity of urban Vehicular Ad Hoc networks (VANETs). The aforementioned
work all assumed that the number of vehicles or vehicular density
is sufficiently large and utilized asymptotic analysis to study the
capacity scaling law, which is only applicable when the number of
vehicles or vehicular density is sufficiently large. In our previous
work \cite{CHEN16}, we considered a vehicular network scenario where
there is only one vehicle, termed vehicle-of-interest (VoI), with
download request from the Internet and all other vehicles cooperate
to assist the communication of the VoI, and analyzed the achievable
throughput of the VoI assuming a cooperative communication strategy.
The scenario being studied correspond to a sparse network scenario
where there are a very small number of vehicles with download requests.

In this paper, we consider a typical delay-tolerant application scenario
with a subset of vehicles, termed Vehicles of Interest (VoIs), having
download requests. Each VoI downloads a distinct large-size file from
the Internet and other vehicles without download requests, termed
\textit{\textcolor{black}{helpers}}, assist the delivery of the files
to the VoIs. A cooperative communication strategy is proposed that
explores the combined use of V2I communications, V2V communications,
mobility of vehicles and cooperation among vehicles and infrastructure
to improve the capacity of vehicular network. An analytical framework
is developed to model the data dissemination process using this strategy,
and a closed form expression of the capacity is obtained, which reveals
the relationship between the capacity and its major performance-impacting
parameters such as inter-infrastructure distance, radio ranges of
infrastructure and vehicles, sensing range of vehicles, transmission
rates of V2I and V2V communications, vehicular density and the proportion
of VoIs. Numerical result shows that the proposed cooperative communication
strategy significantly boosts the capacity of vehicular networks,
especially when the proportion of VoIs is low. Different from the
single-VoI scenario studied in \cite{CHEN16}, when there are multiple
vehicles with download requests, the possible contention and collision
among vehicles in vehicular communications become both important and
challenging issue to study. Furthermore, the work presented in this
paper distinguishes from previous work \cite{Wang14,Huang2015} in
that we focus on an accurate analysis (versus asymptotic analysis)
of the capacity of vehicular networks with a moderate vehicular density
and explore the combined use of V2I communications, V2V communications,
mobility of vehicles and cooperation among vehicles and infrastructure
to improve the achievable capacity of vehicular network, whereas the
results obtained in previous work \cite{Wang14,Huang2015} are only
applicable when the number of vehicles or vehicular density is very
large.

Specifically, the following contributions are made in the paper: 
\begin{enumerate}
\item We propose a novel cooperative communication strategy, which utilizes
V2I communications, V2V communications, mobility of vehicles, and
cooperation among vehicles and infrastructure to boost capacity of
vehicular networks;
\item We develop an analytical framework to model and investigate the data
dissemination process assuming the aforementioned cooperative communication
strategy, and a closed-form expression of the capacity achieved by
the VoIs in a vehicular network with a finite vehicular density is
obtained, which reveals the relationship between the capacity and
its major performance-impacting parameters;
\item Both simulations and numerical analysis are conducted, which show
that the proposed cooperative strategy significantly improves the
capacity of vehicular networks, compared with its non-cooperative
counterpart, even when the proportion of VoIs is small. 
\end{enumerate}
Our results shed light on the optimum deployment of vehicular network
infrastructure in terms of their interval distance, and the optimum
design of cooperative communication strategy to improve the capacity
of vehicular networks. 

The rest of this paper is organized as follows: Section \ref{sec:Related-Work}
reviews related work. Section \ref{sec:System-Model-and} introduces
the system model, the proposed cooperative communication strategy
and the problem formation. Theoretical analysis of the V2I communications,
V2V communications and the capacity are provided in Section \ref{sec:Theoretical-Analysis }.
In Section \ref{sec:Simulation-and-Discussion}, we validate the analytical
result using simulations and conduct further numerical analysis to
discuss our result and its insight. Section \ref{sec:Conclusion-and-Future}
concludes this paper.

\section{Related Work\label{sec:Related-Work}}

Extensive work in the literature investigated the performance of vehicular
networks, measured by the information propagation speed \cite{Zhang2014},
transmission delay \cite{Du15,Zhu15}, downloaded data volume \cite{Haibo14},
packet reception rate \cite{Das2016}, etc. Among the major techniques
to enhance these performance measures, cooperative communications,
including cooperation among vehicles\textcolor{blue}{{} }\textcolor{black}{\cite{Zhu15,Haibo14,Das2016,Liu2016}},
cooperation among infrastructure points \cite{Li2014,Zhang2013,Kim15,Mauri16},
and cooperation among both vehicles and infrastructure points \cite{Mershad12,Si16,Wang16},
stands out as a popular and important technique. In the following,
we review work closely related to the work in this paper. 

The following work investigated cooperative communications among vehicles
in vehicular networks. In \cite{Zhu15}, Zhu et al. studied using
multiple nearby vehicles to collaboratively download data from a RSU
and analyzed the average download time using network coding techniques.
In \cite{Haibo14}, Zhou et al. introduced a cooperative communication
strategy using a cluster of vehicles on the highway to cooperatively
download the same file from the infrastructure to enhance the probability
of successful download. In \cite{Liu2016}, Liu et al. proposed a
centralized cooperative data dissemination scheduling scheme that
utilizes the location information of each vehicle and cooperation
among vehicles to maximize the number of vehicles that successfully
retrieve their requested data in vehicular networks. In \cite{Das2016},
Das et al. introduced a coalitional graph game to model cooperative
message sharing among vehicles in vehicular networks and proposed
a coalition formation algorithm to improve the efficiency of the network
in terms of improving the packet reception rate and reducing transmission
delay. 

Cooperation among infrastructure points can be achieved by caching
different files or different parts of a file into different infrastructure
points to help moving vehicles download from the Internet. In \cite{Zhang2013},
to fully utilize the bandwidth provided by APs, Zhang and Yeo proposed
a cooperative content distribution strategy for vehicles by prefetching
different data into some selected APs, so that vehicles can obtain
the complete data from those selected APs when traveling through their
coverage areas. In \cite{Li2014}, Li et al. proposed a heuristic
content distribution algorithm that caches data in different infrastructure
points by taking data delay and each infrastructure point's storage
limit into account, to maximize the downloaded data size. In \cite{Kim15}
and \cite{Mauri16}, the authors proposed a cooperative content dissemination
scheme utilizing cooperative infrastructure points in vehicular networks
to maximize the success probability of download, utilizing greedy
algorithm and integer linear programming optimization respectively. 

Studies considering both vehicular cooperation and infrastructure
cooperation are comparatively scarce. By exploring cooperation among
vehicles and inter-connected infrastructure points, Mershad et al.
\cite{Mershad12} designed an optimum routing algorithm to reduce
end-to-end delay for delivering a packet from a source to its destination;
and Si et al. \cite{Si16} designed an optimum distributed data hopping
mechanism to enable delay-tolerant data routing over a vehicular network.
In \cite{Wang16}, Wang et al. proposed a scheme that utilizes moving
vehicles to serve as relays to assist data dissemination to a target
vehicle, and the relay selection was conducted by the cooperative
infrastructure points. They focused on reducing the transmission outage
of the target vehicle. 

In this paper, we propose a cooperative communication strategy that
explores the combined use of V2I communications, V2V communications,
mobility of vehicles and cooperation among vehicles and infrastructure
to improve the capacity of vehicular network, and analyze the data
dissemination process and the capacity of the network.

\section{System Model and Problem Formation\label{sec:System-Model-and}}

In this section, we introduce the system model and assumptions used
in the analysis, and also give a rigorous definition of the problem
studied in the paper. 

Specially, we consider a scenario where some VoIs (with proportion
$0<p<1$) want to download large files, e.g., videos, from a remote
server and the file to be downloaded by different VoI is different. 

\subsection{Network Model}

We consider a bi-directional highway segment with length $L$ where
roadside infrastructure, e.g., RSUs, Wi-Fi APs or LTE base stations,
are uniformly deployed along the highway and separated by equal distance
$d,d\ll L$. The width of a lane is typically small compared with
the transmission range of vehicles. Therefore, we ignore the road
width and model multiple lanes in the same direction as one lane \cite{Abboud14,Zhang12,Wis2007}.
We further assume that all infrastructure points are connected to
the Internet through wired or wireless backbone with much larger capacity
than the vehicular network.

\begin{figure}
\centering{}\includegraphics[width=7.6cm]{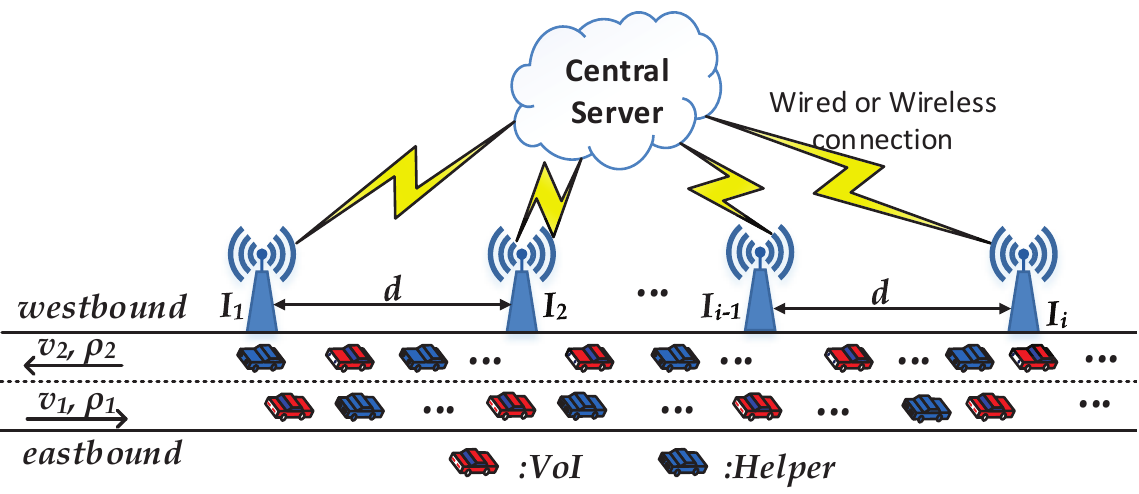} \caption{An illustration of the system model for a bi-directional highway with
infrastructure regularly deployed with equal distance $d$. The density
and speed of vehicles in each direction are $\rho_{1}$, $v_{1}$
and $\rho_{2}$, $v_{2}$ respectively.\label{Fig: System model} }
\end{figure}

We adopt a widely used traffic model in highway \cite{Wis2007,Reis14,Ruixue13}
that the distribution of eastbound and westbound vehicles follows
a homogeneous Poisson process with densities $\rho_{1}$ and $\rho_{2}$
respectively. It follows that the inter-vehicle distances in each
direction are exponentially distributed. This exponential inter-vehicle
spacing distribution has been supported by some empirical study that
it can accurately characterize real traffic distribution when the
traffic density is low or medium \cite{Wis2007}. Furthermore, as
a ready consequence of the superposition property of Poisson processes
\cite{Nelson95}, all vehicles on the highway are also Poissonly distributed
with density $\rho=\rho_{1}+\rho_{2}$. We assume that the proportion
of VoIs travel towards each direction is $p$ ($0<p<1)$. Therefore,
VoIs and \textit{\textcolor{black}{\emph{helpers}}}\emph{ }respectively
have traffic density $p\rho$ and $(1-p)\rho$. Moreover, we assume
that eastbound and westbound vehicles travel at a constant speed of
$v_{1}$ and $v_{2}$ respectively. In reality, individual vehicular
speed may deviate from the mean speed. However, we will show later
that our analysis also applies to other time-varying speed model,
e.g., Gaussian speed model \cite{Zhang2014,Ge15}. The system model
is illustrated in Fig. \ref{Fig: System model}.

\subsection{Wireless Communication Model}

Both V2I and V2V communications are considered. All infrastructure
points are assumed to have the same radio range, denoted by $r_{I}$;
and all vehicles are assumed to have the same radio range, denoted
by $r_{0}$. A pair of vehicles (or vehicle and infrastructure) can
directly communicate with each other if and only if (iff) their Euclidean
distance is not larger than the radio range $r_{0}$ (or $r_{I}$).
\textcolor{black}{There are other more realistic and intricate connection
models, e.g., the SINR connection model \cite{Du15} and the log-normal
connection model \cite{Mao2006}.}\textcolor{blue}{{} }This simplified
unit disk model has been extensively used in the field \cite{Zhang12,Reis14,Mao09}.
It grossly captures the fact that all wireless devices have a limited
transmission range and that the closer two devices become, the easier
it is for them to establish a connection. 

\textcolor{black}{We consider that each vehicle has a single antenna
so that they cannot transmit and receive at the same time. Besides,
we consider a unicast scenario and assume that each infrastructure
(or vehicle) can only transmit information to one vehicle at a time.
Both broadcast and unicast are important in vehicular network \cite{Cheng15}.
For some safety-related applications, e.g., disseminating a message
about an accident on the road, it is better to use broadcast to inform
as many vehicles as possible. Unicast is particularly important when
disseminating delay-tolerant data. Furthermore, it has been shown
in \cite{Gupta} that whether the infrastructure (or vehicle) transmit
to one vehicle at a time, or divides its bandwidth among multiple
users and transmits to multiple users at the same time, does not affect
capacity calculation. }

\textcolor{black}{We further assume that V2I and V2V communications
are allocated different channels so that there is no mutual interference
between them. For V2V communications, CSMA media access control (MAC)
protocol is adopted with sensing range $R_{c}$, which also resembles
the widely used Protocol Interference Model \cite{Kumar04} in wireless
networks. Moreover, we assume V2I and V2V communicate at a constant
data rate $w_{I}$ and $w_{V}$ respectively \cite{Gupta,Mao13,Yang15}.
This simplification allows us to omit physical layer details and focus
on the topological impact of vehicular networks on the capacity, which
is the main performance determining factor. For time-varying channels,
the values of $w_{I}$ and $w_{V}$ can be replaced by the respective
time-averaged data rate of V2I and V2V communications and our analysis
still applies. }

We consider a prioritized V2I transmitting scheme that infrastructure
will transmit its data to VoIs first, i.e., \textit{\textcolor{black}{\emph{helpers}}}
can receive data from infrastructure only when there is no VoI within
the coverage of infrastructure. For V2V communications, \textit{\textcolor{black}{\emph{helpers}}}
function as transmitters and VoIs as receivers. A transmitter can
choose a receiver from either direction within its transmission range.
We limit both V2I and V2V communications to one-hop.\textcolor{blue}{{}
}\textcolor{black}{This can be explained by the fact that in the specific
scenario being considered, there are only a subset of vehicles with
download request (VoIs), all other vehicles (}\textit{\textcolor{black}{\emph{helpers}}}\textcolor{black}{)
assist the VoIs to receive more data. Any new data in the vehicular
network must come from the infrastructure. Therefore, allowing multi-hop
V2V communications between the VoIs and }\textit{\textcolor{black}{\emph{helpers}}}\textcolor{black}{,
e.g., allowing V2V communications between }\textit{\textcolor{black}{\emph{helpers}}}\textcolor{black}{,
only helps to balance the distribution of information stored in }\textit{\textcolor{black}{\emph{helpers}}}\textcolor{black}{{}
but do not increase the net amount of information available in the
network. Furthermore, even though allowing more than one hop V2I communications
between the VoIs and infrastructure is beneficial to the VoI's data
downloading because it allows the VoIs to have longer connection time
(via some intermediate vehicles) with the infrastructure, the improvement
is expected to be marginal, which has been verified by our simulation
result as shown later. }

\subsection{Cooperative Communication Strategy\label{subsec:Cooperative-Communication-Strategy}}

Now we introduce the cooperative communication strategy considered
in this paper. As mentioned previously, we consider a scenario where
some VoIs (with proportion $0<p<1$) want to download large files,
e.g., videos, from a remote server and the file to be downloaded by
different VoI is different. Each requested large file by the VoIs
may be first split into multiple pieces and transmitted to different
infrastructure points such that each infrastructure point has a different
piece of that file, which enables cooperation among infrastructure.
Each piece of data delivered to infrastructure may be further split
and transmitted either directly to the corresponding VoI requesting
it or \textit{\emph{to }}\textit{\textcolor{black}{\emph{helpers}}}
when they move into its coverage so that VoIs and \textit{\textcolor{black}{\emph{helpers}}}
have different pieces of data. Each \textit{\textcolor{black}{\emph{helper}}}
may store data for different VoIs. We assume there is a central server
that has full knowledge of the data transmission process to guarantee
that the data the \textit{\textcolor{black}{\emph{helpers}}} receive
from infrastructure is the data required by the VoIs they will encounter.
This assumption helps to establish the maximum capacity that can be
achieved. Therefore, when the VoIs are in the coverage of infrastructure,
they receive data directly from the infrastructure. In the meantime,
the \textit{\textcolor{black}{\emph{helpers}}} may also receive different
pieces of data from the infrastructure when they obtain access to
the infrastructure. When the VoIs move outside the coverage of infrastructure,
they may continue to receive data from \textit{\textcolor{black}{\emph{helpers}}},
exploiting the mobility of vehicles and V2V communications. In this
way, V2I communications between the VoIs and infrastructure, between
\textit{\textcolor{black}{\emph{helpers}}} and infrastructure, V2V
communications between the VoIs and \textit{\textcolor{black}{\emph{helpers}}},
cooperations among infrastructure and among vehicles, as well as vehicular
mobility are coherently combined to maximize the capacity of the VoIs.
Furthermore, we consider that some practical issues like out of sequence
data delivery can be handled by techniques such as network coding
(e.g., our previous paper \cite{WangPeng15}) so that we can focus
on the main theme of the paper without the need for considering their
impacts. 

\subsection{Problem Formation\label{subsec:Problem-Formation}}

Now we give a formal definition of the capacity considered in this
paper. Consider an arbitrarily chosen time interval $[0,t]$ and denote
the amount of data received by all VoIs as $D(t)$ during this time
interval, which includes data received both directly from infrastructure
and indirectly from \textit{\textcolor{black}{\emph{helpers}}}. In
this paper, we are interested in finding the long-term achievable
capacity of the VoIs using our cooperative communication strategy,
where the long-term capacity, denoted by $\eta$, is formally defined
as follows:
\begin{equation}
\eta=\lim_{t\rightarrow\infty}\frac{D(t)}{t}.\label{eq:definition of capacity}
\end{equation}

\section{Analysis of the Capacity \label{sec:Theoretical-Analysis }}

In this section, we will give detailed analysis of the achievable
capacity by the VoIs, including analyzing the capacity achieved directly
from infrastructure through V2I communications and the capacity achieved
indirectly from \textit{\textcolor{black}{\emph{helpers}}} through
V2V communications. 

\begin{figure}
\centering{}\includegraphics[width=8cm]{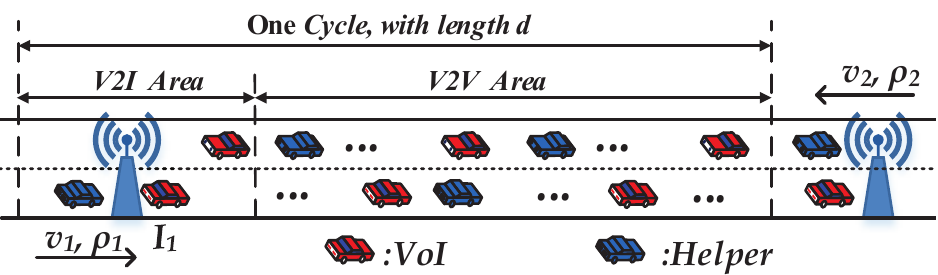} \caption{An illustration of one \textit{\textcolor{black}{cycle}}, which includes
\textit{V2I Area} and \textit{V2V Area}. \label{Cycle} }
\end{figure}

We define the area covered by one infrastructure point (termed \textit{V2I
Area}) and the adjacent area between two consecutive infrastructure
points but not covered by the infrastructure point (termed \textit{V2V
Area}) as a \textit{cycle}, which has length $d$. See Fig. \ref{Cycle}
for an illustration. It follows from the renewal theory \cite{Gallager13}
that the long-term achievable capacity by the VoIs from each cycle,
denoted by $\eta_{\text{cycle}}$, is identical and the total capacity
achieved in a given highway segment with length $L\gg d$ can be readily
calculated by (ignoring the trivial fact that $\frac{L}{d}$ may not
be an integer):
\begin{equation}
\eta=\frac{L}{d}\eta_{\text{cycle}}.\label{eq:relation between total capacity and capacity from one cycle}
\end{equation}

From \eqref{eq:relation between total capacity and capacity from one cycle},
to calculate the total capacity achievable by the VoIs from a highway
segment with length $L$, it suffices to calculate the capacity achieved
by the VoIs from one cycle, which includes capacity achieved both
from V2I communications and V2V communications, given as follows:
\begin{equation}
\eta_{\text{cycle}}=\lim_{t\rightarrow\infty}\frac{D_{V2I}(t)}{t}+\lim_{t\rightarrow\infty}\frac{D_{V2V}(t)}{t}\label{eq:definition of capcity achievable from one cycle}
\end{equation}
where $D_{V2I}(t)$ and $D_{V2V}(t)$ are respectively the expected
amount of data received by the VoIs from infrastructure in the V2I
Area and from \textit{\textcolor{black}{\emph{helpers}}} in the V2V
Area during time period $t$. In the following, we will focus on studying
one cycle entirely contained within the highway segment of length
$L$, termed the \emph{cycle of interest}. We will first calculate
the two terms on the right hand side of  \eqref{eq:definition of capcity achievable from one cycle}
separately, and then combine both terms to obtain the final expression
of the achievable capacity. 

\subsection{Capacity achieved by VoIs from V2I communications \label{subsec:Theoretical-Analysis-of of V2I communications}}

Without loss of generality, we call the infrastructure point located
in our cycle of interest $I_{1}$. We assume that time is divided
into time slots of equal length $\triangle t$, and $\triangle t$
is sufficiently small that we can approximately regard vehicles as
stationary during each time slot. Denote by $q_{1}(i),i=1,2,...$
a discrete random variable representing the fraction of time that
VoIs' V2I communication with $I_{1}$ happens during the $i$th time
slot $[(i-1)\triangle t,i\triangle t)$. Recall that infrastructure
delivers its data directly to the VoIs as long as there are VoIs within
its coverage. Therefore, $q_{1}(i)$ is equal to 1 when there exist
at least one VoIs within the coverage of $I_{1}$ during the $i$th
time slot; otherwise it is equal to 0. This follows that the expected
total amount of data the VoIs can obtain from one cycle through V2I
communications during time period $[0,t]$ can be calculated by (ignoring
the trivial fact that $\frac{t}{\triangle t}$ may not be an integer):
\begin{equation}
D_{V2I}(t)=\lim_{\triangle t\rightarrow0}E\left(w_{I}\sum_{i=1}^{t/\triangle t}q_{1}(i)\triangle t\right).\label{eq:data achieved from V2I communication by the VoIs}
\end{equation}

According to the ergodicity and stationarity properties of homogeneous
Poisson point process \cite{Baccelli09}, the time average of $q_{1}(i)$
is equal to the probability that there is at least one VoI within
the coverage of $I_{1}$ at a randomly chosen time slot, denoted by
$\bar{q}_{1}$. Considering the Poisson distribution of vehicles,
it can be readily shown that $\bar{q}_{1}=1-e^{-p\rho2r_{I}}$. Therefore,
we have
\begin{align}
\lim_{t\rightarrow\infty}\frac{\lim_{\triangle t\rightarrow0}E\left(\sum_{i=1}^{t/\triangle t}q_{1}(i)\triangle t\right)}{t} & =\bar{q}_{1}=1-e^{-p\rho2r_{I}}.\label{eq:probability that there are at least one VoI within the coverage of I_1}
\end{align}
Combing \eqref{eq:data achieved from V2I communication by the VoIs}-\eqref{eq:probability that there are at least one VoI within the coverage of I_1},
we have the long-term capacity achieved by the VoIs from one cycle
through their V2I communications:
\begin{equation}
\lim_{t\rightarrow\infty}\frac{D_{V2I}(t)}{t}=w_{I}\left(1-e^{-p\rho2r_{I}}\right).\label{eq: capacity achieved by the VoIs through V2I communications from one cycle}
\end{equation}

\subsection{Capacity achieved by VoIs from V2V communications\label{subsec:Theoretical-Analysis-of V2V communications}}

Note that the data received by the VoIs from \textit{\textcolor{black}{\emph{helpers}}}
through V2V communications eventually comes from the data received
by the \textit{\textcolor{black}{\emph{helpers}}} from infrastructure
during their V2I communications. Therefore, the amount of data the
VoIs can receive from V2V communications during time period $[0,t]$,
on one hand, is constrained by how much data the \textit{\textcolor{black}{\emph{helpers}}}
can receive via their V2I communications during time period $[0,t]$;
on the other hand, is limited by how much data the \textit{\textcolor{black}{\emph{helpers}}}
can transmit to the VoIs through V2V communications during time period
$t$. Taking the above two constraints into account, we have the following
results:
\begin{thm}
\label{thm:the capacity achieved from V2V communiactions by the VoIs}The
capacity the VoIs can achieve through V2V communications from one
cycle is given by:
\begin{equation}
\lim_{t\rightarrow\infty}\frac{D_{V2V}(t)}{t}=\min\left\{ \lim_{t\rightarrow\infty}\frac{D_{I_{-}H}(t)}{t},\lim_{t\rightarrow\infty}\frac{D_{V}(t)}{t}\right\} ,\label{eq:definition of the capacity achieved from V2V communications by the VoIs}
\end{equation}
where $D_{I_{-}H}(t)$ is the expected amount of data received by
helpers from one cycle through their V2I communications during time
period $[0,t]$, and $D_{V}(t)$ is the expected amount of data the
helpers can deliver to the VoIs through V2V communications in the
V2V Area during time period $[0,t]$ without considering the limitation
of the amount of data received by helpers from the infrastructure. 
\end{thm}
Since the bottleneck is either in the V2V communications between the
VoIs and \textit{\textcolor{black}{\emph{helpers}}}, represented by
$\lim_{t\rightarrow\infty}\frac{D_{V}(t)}{t}$, or in the V2I communications
between the \textit{\textcolor{black}{\emph{helpers}}} and the infrastructure,
represented by $\lim_{t\rightarrow\infty}\frac{D_{I_{-}H}(t)}{t}$,
the proof of Theorem \ref{thm:the capacity achieved from V2V communiactions by the VoIs}
follows readily. More specifically, imagine the V2V communication
process between \textit{\textcolor{black}{\emph{helpers}}} and the
VoIs as a single-queue queuing system. The rate the \textit{\textcolor{black}{\emph{helpers}}}
receive data from the infrastructure, $\lim_{t\rightarrow\infty}\frac{D_{I_{-}H}(t)}{t}$,
is equivalent to the incoming rate of the queue. The rate the \textit{\textcolor{black}{\emph{helpers}}}
deliver data to the VoIs, $\lim_{t\rightarrow\infty}\frac{D_{V}(t)}{t}$,
is equivalent to the processing speed of the queue. The outgoing rate
of the queue, $\lim_{t\rightarrow\infty}\frac{D_{V2V}(t)}{t}$, is
either equal to the incoming rate or equal to the processing speed.

From Theorem \ref{thm:the capacity achieved from V2V communiactions by the VoIs},
to obtain the capacity achieved by the VoIs through V2V communications
with the \textit{\textcolor{black}{\emph{helpers}}} from one cycle,
it remains to calculate the long-term capacity achieved by the \textit{\textcolor{black}{\emph{helpers}}}
from one cycle through their V2I communications, $\lim_{t\rightarrow\infty}\frac{D_{I_{-}H}(t)}{t}$,
and the long-term capacity the VoIs can achieve through V2V communications
from one cycle without considering the limitation of the amount of
data the \textit{\textcolor{black}{\emph{helpers}}} received, $\lim_{t\rightarrow\infty}\frac{D_{V}(t)}{t}$.
In the following, we will calculate these two terms separately. 

\subsubsection{Calculation of $\lim_{t\rightarrow\infty}\frac{D_{I_{-}H}(t)}{t}$}

Denote by $q_{2}(i)$ a discrete random variable, which is equal to
1 when \textit{\textcolor{black}{\emph{helpers}}}\textit{\textcolor{black}{'}}
V2I communication happens during the $i$-th time slot $[(i-1)\triangle t,i\triangle t)$,
$i=1,2,...$, otherwise it is equal to 0. Similar to the analysis
in section \ref{subsec:Theoretical-Analysis-of of V2I communications},
we have
\begin{align}
\lim_{t\rightarrow\infty}\frac{D_{I_{-}H}(t)}{t} & =\lim_{t\rightarrow\infty}\frac{w_{I}\lim_{\triangle t\rightarrow0}E\left(\sum_{i=1}^{t/\triangle t}q_{2}(i)\triangle t\right)}{t}\nonumber \\
 & =w_{I}\bar{q}_{2},\label{eq:capacity  the helpers achieve from one cycle_cal}
\end{align}
where $\bar{q}_{2}$ is the probability that \textit{\textcolor{black}{\emph{helpers}}}\textit{\textcolor{black}{'}}
 V2I communication happens at a randomly chosen time slot. Note that
an infrastructure point only delivers its data to \textit{\textcolor{black}{\emph{helpers}}}
when both of the following conditions are met: (i) there is no VoI
within its coverage, \textit{and} (ii) there is at least one \textit{\textcolor{black}{\emph{helpers}}}\emph{
}\textit{\emph{within its coverage}}\emph{.} Thus, $\bar{q}_{2}$
can be readily calculated by using the Poisson distribution of the
VoIs and the \textit{\textcolor{black}{\emph{helpers}}}:
\begin{equation}
\bar{q}_{2}=e^{-p\rho2r_{I}}\left(1-e^{-(1-p)\rho2r_{I}}\right)=e^{-p\rho2r_{I}}-e^{-\rho2r_{I}}.\label{eq:probability that the infrastructure delivers its data to the helper in a time slot}
\end{equation}
Combing \eqref{eq:capacity  the helpers achieve from one cycle_cal}
and \eqref{eq:probability that the infrastructure delivers its data to the helper in a time slot},
we have:
\begin{equation}
\lim_{t\rightarrow\infty}\frac{D_{I_{-}H}(t)}{t}=w_{I}\bar{q}_{2}=w_{I}\left(e^{-p\rho2r_{I}}-e^{-\rho2r_{I}}\right).\label{eq:capacity  the helpers achieve from one cycle}
\end{equation}

\subsubsection{Calculation of $\lim_{t\rightarrow\infty}\frac{D_{V}(t)}{t}$}

In this subsection, we analyze the maximum capacity achieved by the
VoIs through V2V communications from one cycle\emph{ }area without
considering the amount of data each \textit{\textcolor{black}{\emph{helper}}}
has. 

Recall that for V2V communications, we adopt CSMA multiple access
protocol with sensing range $R_{c}$. Therefore, a \textit{\textcolor{black}{\emph{helper}}}
within the \textcolor{black}{V2V Area} can potentially be chosen as
one of the simultaneously transmitters when there is no other \textit{\textcolor{black}{\emph{helper}}}
transmitting within its sensing range\textit{ and}\textit{\emph{ there
is at least one VoIs within its transmission range}}\textit{.} We
call a \textit{\textcolor{black}{\emph{helper}}}\textit{\emph{,}}\emph{
}together with the VoI that the \textit{\textcolor{black}{\emph{helper}}}\emph{
}transmits to, an\textit{\textcolor{black}{{} active helper-VoI pair}}
iff this \textit{\textcolor{black}{\emph{helper}}}\emph{ }is chosen
as a transmitter \textit{\emph{and}} chooses this VoI within its transmission
range as its receiver.

Denote by $N_{p}^{\chi}(i)$ the number of simultaneous \textit{\textcolor{black}{\emph{helper-VoI}}}
pairs in the \textcolor{black}{V2V Area} during the $i$-th time slot
$[(i-1)\triangle t,i\triangle t)$, where the superscript $\chi\in\Phi$
denotes the scheduling algorithm that selects the simultaneously active
\textit{\textcolor{black}{\emph{helper-VoI}}} pairs and $\Phi$ denotes
the set of all scheduling algorithms. Using the analysis in \cite{Mao13},
it can be shown that:
\begin{align}
\lim_{t\rightarrow\infty}\frac{D_{V}^{\chi}(t)}{t} & =\lim_{t\rightarrow\infty}\frac{w_{V}\lim_{\triangle t\rightarrow0}E\left(\sum_{i=1}^{t/\triangle t}N_{p}^{\chi}(i)\triangle t\right)}{t}\nonumber \\
 & =w_{V}E\left[N_{p}^{\chi}\right],\label{eq:calculation of capacity achieved by the VoIs from V2V communications}
\end{align}
where $D_{V}^{\chi}(t)$ is the respective $D_{V}(t)$ assuming the
scheduling scheme $\chi$, and $E\left[N_{p}^{\chi}\right]$ is the
expected number of simultaneously active \textit{\textcolor{black}{\emph{helper-VoI}}}
pairs in the \textit{\emph{V2V Area}} at a randomly chosen time slot. 

From \eqref{eq:calculation of capacity achieved by the VoIs from V2V communications},
the maximum $\lim_{t\rightarrow\infty}\frac{D_{V}^{\chi}(t)}{t}$
is achieved when using an optimum scheduling algorithm that schedules
as many active \textit{\textcolor{black}{\emph{helper-VoI}}} pairs
as possible. Therefore, in the following analysis, we shall establish
the optimum scheduling algorithm and the maximum $E\left[N_{p}^{\chi}\right]$
that can be achieved by the algorithm. Specifically, we will first
find an optimum scheduling scheme, denoted by $\chi_{opt}$, that
leads to the maximum number of simultaneously active \textit{\textcolor{black}{\emph{helper-VoI}}}
pairs, and then calculate $E\left[N_{p}^{\chi_{opt}}\right]$ under
this optimum algorithm. Without loss of generality, we designate the
left boundary point of the V2V Area, i.e., the point to the right
of infrastructure point $I_{1}$ and at a distance $r_{I}$ to $I_{1}$,
as the origin of the coordinate system, and the east (right) direction
as positive ($+x$) direction. The following theorem summarizes the
optimum scheduling scheme.
\begin{thm}
\label{thm:The-optimum-selection scheme}An optimum scheduling scheme
$\chi_{opt}$, which leads to the maximum number of simultaneously
active helper-VoI pairs in V2V Area is as follows: select active helper-VoI
pairs in order from left to the right. First, choose the first helper
to the right of the origin that has at least one VoIs within its coverage
as the first transmitter, and the left-most VoI within the coverage
of that helper as its receiver. The next transmitter is the nearest
helper to the current transmitter, and satisfies the following conditions:
1) the distance between this helper and the current transmitter is
no smaller than $R_{c}$; 2) it can find at least one VoIs within
its coverage, which is different from the receiver of the current
transmitter. If there are multiple VoIs, always chooses the leftmost
VoI. Repeat the above process until the rightmost border of the V2V
Area is reached.
\end{thm}
\begin{IEEEproof}
See Appendix A. 
\end{IEEEproof}
\begin{rem}
Note that the optimum scheduling algorithm that achieves the maximum
number of active \textit{\textcolor{black}{\emph{helper-VoI}}} pairs
may not be unique.
\end{rem}
Now we calculate the maximum expected number of \textit{\textcolor{black}{\emph{helper-VoI}}}
pairs, $E\left[N_{P}^{\chi_{opt}}\right]$, and the corresponding
value of $\lim_{t\rightarrow\infty}\frac{D_{V}(t)}{t}$ under the
optimum scheme $\chi_{opt}$.

\begin{figure}[t]
\centering{}\includegraphics[width=8cm]{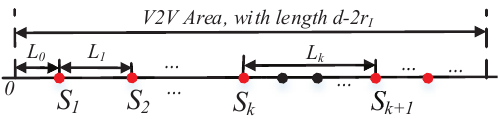} \caption{An illustration of the distribution of distances between two consecutive
simultaneous transmitters.\label{illustration of the distance between two consecutive chosen helpers} }
\end{figure}

Denote by $S_{k}\in[0,d-2r_{I}],k=1,2,...$. the position of the $k$-th
transmitter (\textit{\textcolor{black}{\emph{helper}}} in the active
\textit{\textcolor{black}{\emph{helper-VoI}}} pair) under the optimum
scheduling scheme $\chi_{opt}$. Denote by $L_{k},k=1,2,...$ the
distance between the $k$-th and the $(k+1)$-th transmitter, and
$L_{0}$ the distance between the first transmitter and the origin.
See Fig. \ref{illustration of the distance between two consecutive chosen helpers}
for an illustration. It is straightforward that $L_{k},k=1,2,...$
are identically and independently distributed (i.i.d.) and are also
independent of $L_{0}$. \textcolor{black}{Note that the distribution
of $L_{0}$ is not the same as that of $L_{k},k=1,2,...$. This is
due to the fact that when selecting the first transmitter assuming
$\chi_{opt}$, we directly choose the first (leftmost) helper within
$[0,d-2r_{I}]$ that has at least one VoI within its coverage. In
contrast, when $k>1$, we select the $k$-th transmitter from helpers
located within $[S_{k-1}+R_{c},d-2r_{I}]$, which imposes an additional
condition that $L_{k}>R_{c}$. It can be shown that the expected number
of helper-VoI pairs, $E\left[N_{P}^{\chi_{opt}}\right]$, is exactly
the expected number of renewals of a renewal counting process in the
V2V Area, with a delay length of $E[L_{0}]$ (as $L_{0}$ has a different
distribution from $L_{k},k=1,2,...$) and each renewal has an average
length $E[L_{k}],k=1,2,...$. }Using the renewal theory \cite{Gallager13},
$E\left[N_{p}^{\chi_{opt}}\right]$ can be calculated as follows:
\begin{equation}
E\left[N_{p}^{\chi_{opt}}\right]=\sum_{n=1}^{\infty}Pr(\sum_{i=0}^{n-1}L_{i}\leq d-2r_{I}),\label{eq:Exact calculation of expected number of simultaneous pairs}
\end{equation}

An alternative way of obtaining  \eqref{eq:Exact calculation of expected number of simultaneous pairs}
is by noting that the inner term of  \eqref{eq:Exact calculation of expected number of simultaneous pairs},
i.e., $Pr(\sum_{i=0}^{n-1}L_{i}\leq d-2r_{I})$, gives the cumulative
distribution function (cdf) of the maximum number of simultaneously
active \textit{\textcolor{black}{\emph{helper-VoI}}} pairs assuming
the optimum scheduling $\chi_{opt}$. It then follows that the summation
of the cdf gives the expected value of $N_{p}^{\chi_{opt}}$. 

Equation \eqref{eq:Exact calculation of expected number of simultaneous pairs}
shows that to calculate $E\left[N_{p}^{\chi_{opt}}\right]$, we first
need to calculate the distribution of each $L_{k},k=0,1,2,...$. The
following two theorems characterize the probability density function
(\textit{\emph{pdf}}) of $L_{k}$, denoted by $f_{L_{k}}(x),k=1,2..$.
and the \textit{\emph{pdf}} of $L_{0}$, denoted by $f_{L_{0}}(x)$
respectively. 
\begin{thm}
\label{thm:Distribution of Lk}Consider a bi-directional vehicular
network with vehicular densities $\rho_{1}$ and $\rho_{2}$, with
$p$ percentage of vehicles being the VoIs and the remaining $1-p$
percentage of vehicles being the helpers. Furthermore, each vehicles'
radio range is $r_{0}$ and sensing range is $R_{c}$. The distance
between two consecutive transmitters (helpers in two consecutive active
helper-VoI pairs), $L_{k},k\geq1$, under the optimum scheduling scheme
$\chi_{opt}$, has the pdf as follows:
\begin{align}
 & f_{L_{k}}(x)\nonumber \\
= & \begin{cases}
\sum_{m=1}^{\infty}f(x-R_{c};m,(1-p)\rho)Pr(m_{k}=m), & x\geq R_{c}\\
0 & x<R_{c}
\end{cases},\label{eq:pdf of Lk}
\end{align}
where $f(x;k,\alpha)=\frac{\alpha^{k}x^{k-1}e^{-\alpha x}}{(k-1)!}$
is the pdf of Erlang distribution with shape parameter $k$ and rate
parameter $\alpha$, and \textup{\textcolor{black}{$m_{k}$}} is the
(random) number of helpers within\textup{\textcolor{black}{{} $\left[S_{k}+R_{c},\min\left\{ S_{k+1},d-2r_{I}\right\} \right)$.
}}\textcolor{black}{The probability mass function of $m_{k}$ is given
by: }
\begin{equation}
Pr(m_{k}=1)\thickapprox1-e^{-p\rho2r_{0}},\label{eq:probability that mk equals to 1}
\end{equation}
and for $m\geq2$,
\begin{equation}
\begin{aligned} & Pr(m_{k}=m)\\
\approx & e^{-p\rho2r_{0}}\left(p-pe^{-\rho2r_{0}}\right)\left(1-p+pe^{-\rho2r_{0}}\right)^{m-2}.
\end{aligned}
\label{eq:probability that mk equals to m}
\end{equation}
Moreover, when $R_{c}\geq2r_{0}$, the approximation in \textup{\eqref{eq:probability that mk equals to 1}}
and \textup{\eqref{eq:probability that mk equals to m}} becomes accurate
and can be replaced by equality.\textup{ }
\end{thm}
\begin{IEEEproof}
See Appendix B.
\end{IEEEproof}
\begin{thm}
\label{thm:Distribtution of L0}Under the same setting as that described
in Theorem \ref{thm:Distribution of Lk}, the distance between the
first transmitter and the origin under the optimum scheduling scheme
$\chi_{opt}$, $L_{0}$, has the pdf as follows:
\begin{equation}
f_{L_{0}}(x)=\sum_{m=1}^{\infty}f\left(x;m,(1-p)\rho\right)Pr(m_{0}=m),\label{eq:pdf of L0}
\end{equation}
where\textcolor{blue}{{} }\textup{\textcolor{black}{$m_{0}$ }}\textcolor{black}{is
the (random) number of helpers within $\left(0,S_{1}\right)$ with
distribution:} $Pr(m_{0}=1)=1-\left(1-p+pe^{-\rho r_{0}}\right)e^{-p\rho r_{0}}$;
and for $m\geq2$,
\[
Pr(m_{0}=m)=e^{-p\rho r_{0}}(1-c_{1})c_{1}^{m-2}\left(1-p+pe^{-\rho r_{0}}\right),
\]
where $c_{1}=1-p+pe^{-\rho2r_{0}}$. 
\end{thm}
\begin{IEEEproof}
The cdf of $L_{0}$ can be derived using the same method as that used
in the proof of Theorem \ref{thm:Distribution of Lk}. Particularly,
we have $L_{0}=\sum_{i=0}^{m_{0}-1}l_{0,i}$, and in this case, define
$h_{0,i}=\min\left\{ l_{0,i},2r_{0}\right\} ,i=1,...m_{0}-1$, $h_{0,0}=\min\left\{ l_{0,0},r_{0}\right\} +r_{0}$,
and define $H_{0,m}=\sum_{i=1}^{m-2}h_{0,i}+h_{0,0}$. See proof of
Theorem \ref{thm:Distribution of Lk} for definitions of these parameters.
The proof follows. 
\end{IEEEproof}
From  \eqref{eq:pdf of Lk} and  \eqref{eq:pdf of L0}, we can see
that the $pdf$ of $L_{0}$ and $L_{k},k=1,2,...$ are both in the
forms of rather complicated expressions. Accordingly, the computation
of the distribution of $\sum_{i=0}^{n-1}L_{i},n=1,2,...$, which is
required for computing $E\left[N_{p}^{\chi_{opt}}\right]$ and relies
on the joint the distribution of $L_{k},n=1,2,...$., can become even
more intricate. In our case, assuming that $d$ is much larger compared
with the distance between two consecutive simultaneous transmitters,
i.e., we have $d\gg E[L_{k}]$. It is then reasonably accurate to
calculate the value of $E\left[N_{p}^{\chi_{opt}}\right]$ \textit{\textcolor{black}{approximately}}
using the Elementary Delayed Renewal Theorem \cite[Theorem 5.8.4]{Gallager13},
shown as follows: 
\begin{equation}
E\left[N_{p}^{\chi_{opt}}\right]\approx\frac{d-2r_{I}}{E[L_{k}]},\label{eq:Approximiation of expected number of simultaneous pairs}
\end{equation}
where only the expected value of $L_{k},k=1,2,...$ is needed. In
the following, we first calculate the expected value of $L_{k},k=1,2,...$,
and then use the obtained result of $E[L_{k}]$ to calculate $E\left[N_{p}^{\chi_{opt}}\right]$. 

According to the \textit{pdf} of $L_{k}$ provided in \eqref{eq:pdf of Lk},
the expectation of $L_{k},k=1,2,...$ can be readily calculated as
follows:
\begin{align}
E[L_{k}] & =E\left[R_{c}+\sum_{i=0}^{m_{k}-1}l_{k,i}\right]\nonumber \\
 & =R_{c}+\sum_{m=1}^{\infty}E\left[\sum_{i=0}^{m-1}l_{k,i}\right]Pr(m_{k}=m)\nonumber \\
 & =R_{c}+\sum_{m=1}^{\infty}\frac{m}{(1-p)\rho}Pr(m_{k}=m)\nonumber \\
 & =R_{c}+\frac{p-pe^{-\rho2r_{0}}+e^{-p\rho2r_{0}}}{(1-p)p\rho\left(1-e^{-\rho2r_{0}}\right)},\label{eq:expectation of Lk}
\end{align}
\textcolor{black}{where the first step results by using \eqref{eq:definition of Lk}
and the second step is obtained by using the total probability theorem;
the third step is obtained due to fact that $\sum_{i=0}^{m-1}l_{k,i}$
is the sum of $m$ i.i.d. exponential random variable with mean $\frac{1}{(1-p)\rho}$,
which is independent of $m_{k}$; and the last step results by plugging
in $Pr(m_{k}=m)$ shown as \eqref{eq:Pr(mk=00003Dm)}. }

Putting \eqref{eq:expectation of Lk} into \eqref{eq:Approximiation of expected number of simultaneous pairs}
and simplifying it, we have 
\begin{align}
 & E\left[N_{p}^{\chi_{opt}}\right]\nonumber \\
= & \frac{(1-p)p\rho(1-e^{-\rho2r_{0}})(d-2r_{I})}{(1-p)p\rho(1-e^{-\rho2r_{0}})R_{c}+p-pe^{-\rho2r_{0}}+e^{-p\rho2r_{0}}}.\label{eq:final result of the expected number of simultaneous pairs}
\end{align}
Combining \eqref{eq:calculation of capacity achieved by the VoIs from V2V communications}
and \eqref{eq:final result of the expected number of simultaneous pairs},
we have:
\begin{align}
 & \lim_{t\rightarrow\infty}\frac{D_{V}(t)}{t}\nonumber \\
= & \frac{w_{V}(1-p)p\rho(1-e^{-\rho2r_{0}})(d-2r_{I})}{(1-p)p\rho(1-e^{-\rho2r_{0}})R_{c}+p-pe^{-\rho2r_{0}}+e^{-p\rho2r_{0}}}.\label{eq:maximum capacity the VoIs can achieve from V2V communications}
\end{align}

\subsection{Achievable capacity }

In this subsection, we first give the final result of the capacity
achieved by the VoIs by combing the results from Section \ref{subsec:Theoretical-Analysis-of of V2I communications}
and \ref{subsec:Theoretical-Analysis-of V2V communications}; then
we analyze the capacity achieved by eastbound and westbound VoIs separately
to demonstrate the relationship between capacity achieved by eastbound
and westbound VoIs and their vehicular density. 

\subsubsection{Total Achievable Capacity}

Combining the capacity achieved by the VoIs from V2I communications
and V2V communications from one cycle with length $d$ shown in Section
\ref{subsec:Theoretical-Analysis-of of V2I communications} and \ref{subsec:Theoretical-Analysis-of V2V communications},
the total capacity achieved by the VoIs from a highway segment with
length $L$ can be readily obtained as follows:
\begin{align}
\eta= & \frac{L}{d}\lim_{t\rightarrow\infty}\frac{D_{V2I}(t)+D_{V2V}(t)}{t}\nonumber \\
= & \frac{L}{d}\min\biggl\{ w_{I}\left(1-e^{-\rho2r_{I}}\right),\nonumber \\
 & \left.w_{I}\left(1-e^{-p\rho2r_{I}}\right)+\frac{w_{V}c_{2}(d-2r_{I})}{c_{2}R_{c}+p-pe^{-\rho2r_{0}}+e^{-p\rho2r_{0}}}\right\} ,\label{eq:final result of the total achievable capacity}
\end{align}
where $c_{2}=(1-p)p\rho\left(1-e^{-\rho2r_{0}}\right)$. 
\begin{rem}
It is interesting to note from \eqref{eq:final result of the total achievable capacity}
that the achievable capacity is irrelevant to the velocity of vehicles,
which appears to be counter-intuitive at the first sight. This can
be explained from the data dissemination process. As  \eqref{eq: capacity achieved by the VoIs through V2I communications from one cycle}
and \eqref{eq:calculation of capacity achieved by the VoIs from V2V communications}
show, both the capacity achieved by the VoIs from infrastructure and
from \textit{\textcolor{black}{\emph{helpers}}} only depend on the
spatial distribution of vehicles. In our system, the vehicles' arrival
follows a Poisson process and the vehicles move at a constant speed.
Therefore, the spatial distribution of the vehicles are both stationary
and ergodic \cite{Baccelli09} (ignoring the finite length of the
road segment $L$). It follows that the capacity that can be achieved
by the VoIs is independent of vehicular velocities. This observation
implies that when vehicles arrive following a Poisson process, our
analysis assuming the constant speed model is also applicable to other
time-varying speed model, e.g., Gaussian speed model \cite{Zhang2014},
as long as the resulting spatial distribution of vehicles is time-invariant,
i.e., stationary.
\end{rem}

\begin{rem}
In the extreme case when all vehicles have download requests, i.e.,
when $p=1$, the achievable capacity from one cycle is $\eta_{cycle}=\eta_{max}=w_{I}\left(1-e^{-\rho2r_{I}}\right)$,
which is exactly the achievable capacity when all vehicles directly
receive data from the infrastructure without cooperative V2V communications.
It can be further established when $p$ is greater than a certain
threshold, cooperative V2V communications between the \textit{\textcolor{black}{\emph{helpers}}}
and the VoIs are of little use in boosting the capacity. This can
be explained by that in the particular scenario considered in the
paper, all new data comes from outside the vehicular network. V2V
communications between the \textit{\textcolor{black}{\emph{helpers}}}
and the VoIs only help to extend the communication range of the VoIs
when there is no VoI in the infrastructure's coverage and balance
data among the VoIs, but cannot increase the net amount of data available
in the vehicular network. Therefore, when the density of the VoIs
is high, the probability that there is no VoI in the infrastructure's
coverage is negligible and thus it is more beneficial for the VoIs
to retrieve data directly from the infrastructure. In this situation,
V2V cooperative communications offer little benefit in boosting the
vehicular network capacity. Unsurprisingly, following the argument
outlined earlier, when the vehicular density is sufficiently large,
even for a small value of $p$, the benefit of V2V cooperative communications
vanishes very quickly. This can be also validated using  \eqref{eq:final result of the total achievable capacity}. 
\end{rem}

\subsubsection{Capacity achieved by eastbound and westbound VoIs }

In this subsection, we analyze the capacity achieved from one cycle\emph{
}by all eastbound and westbound VoIs separately, denoted by $\eta_{e}$
and $\eta_{w}$ respectively. The following theorem summarizes the
results.
\begin{thm}
\label{thm:The-capacity-achieved by VoIs from each direction}The
capacities achieved from one cycle by the eastbound and westbound
VoIs, are proportional to the traffic density of eastbound vehicles
and westbound vehicles respectively, which are given by:
\begin{equation}
\eta_{e}=\eta_{\text{cycle}}\cdot\frac{\rho_{1}}{\rho_{1}+\rho_{2}},\label{eq:acheivable capacity from one cycle by VoIs in eastbound}
\end{equation}
and
\begin{equation}
\eta_{w}=\eta_{\text{cycle}}\cdot\frac{\rho_{2}}{\rho_{1}+\rho_{2}}.\label{eq:acheivable capacity from one cycle by VoIs in westbound}
\end{equation}
\end{thm}
\begin{IEEEproof}
See Appendix C.
\end{IEEEproof}
\begin{rem}
It is interesting to note that the capacities achieved by VoIs traveling
in each direction are strictly proportional to the vehicular densities
of that direction respectively. This can be explained by that when
the vehicular density in a direction increases, there is a higher
chance for the VoIs travel in that direction to communicate with the
infrastructure and to receive more data indirectly from the \textit{\textcolor{black}{\emph{helpers}}}
via V2V communication. As a VoI travels in a direction is statistically
indistinguishable from another VoI travels in the same direction,
this result suggests that the achievable throughput by a (any) VoI,
no matter which direction the VoI is traveling in, will be statistically
the same. 
\end{rem}

\section{Simulation and Discussion\label{sec:Simulation-and-Discussion}}

In this section we conduct Monte-Carlo simulations to establish the
accuracy of the theoretical analysis and discuss its insights. Specifically,
we set the length of a highway segment $L$=100km. Eastbound and westbound
vehicles move at constant speeds of $v_{1}$=20m/s and $v_{2}$=25m/s
respectively. The radio ranges of infrastructure points and vehicles
are 400m and 200m (typical radio ranges using DSRC \cite{Haibo14})
respectively. The data rates of V2I and V2V communications are $w_{I}$=20Mb/s
and $w_{V}$=2Mb/s respectively. Each simulation is repeated 2000
times and the average value is shown in the plot.

\begin{figure}[t]
\centering{}\includegraphics[width=8cm]{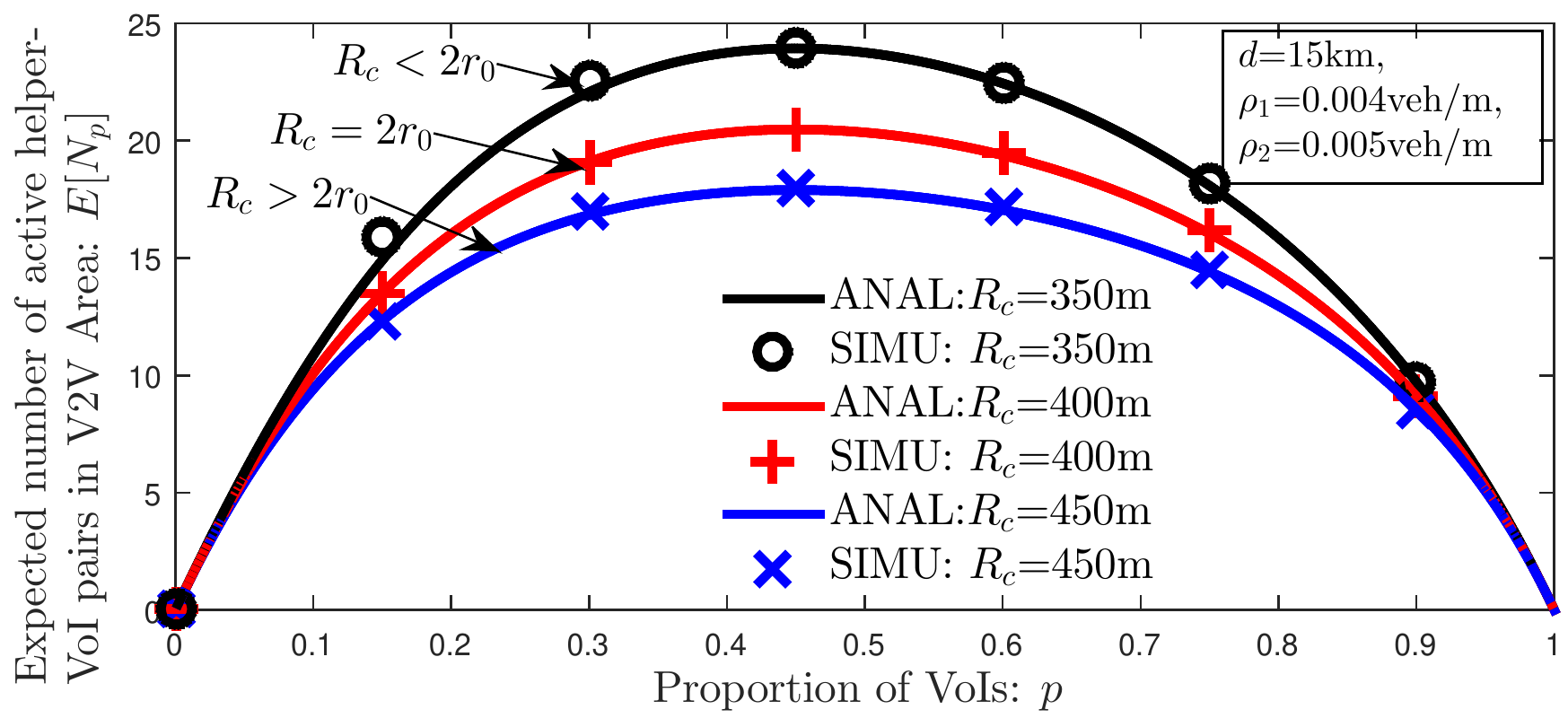}\caption{A comparison of the expected number of simultaneously active \textit{\textcolor{black}{\emph{helper-VoI}}}
pairs in one V2V Area with respect to the proportion of VoIs $p$
between simulation and analysis, for different sensing ranges $R_{c}$. }
\label{SIMU-ANAL-ENP}
\end{figure}

Fig. \ref{SIMU-ANAL-ENP} shows a comparison between the expected
number of simultaneous active \textit{\textcolor{black}{\emph{helper-VoI}}}
pairs in a V2V Area from analysis and simulation, under three different
sensing ranges $R_{c}$: $R_{c}<2r_{0}$, $R_{c}=2r_{0}$ and $R_{c}>2r_{0}$.
It is shown that the analytical result match perfectly with the simulation
result when $R_{c}\geq2r_{0}$. When $R_{c}<2r_{0}$, there is a marginal
gap between the simulation and analytical result, and the gap reduces
with an increase of $p$. This is due to the fact that when $R_{c}<2r_{0}$,
other things being equal, with an increase of $p$, the density of
\textit{\textcolor{black}{\emph{helpers}}} becomes smaller. Therefore,
it is less likely to occur the scenario discussed in the proof of
Theorem \ref{thm:Distribution of Lk} that the VoI of the $k$-th
($k\geq1)$ \textit{\textcolor{black}{\emph{helper-VoI}}} pair is
located within the coverage of the \textit{\textcolor{black}{\emph{helper}}}
$V_{k,1}$, \textit{\textcolor{black}{\emph{helper}}} $V_{k,2},\cdots$.
Consequently, the \textit{\emph{approximations}} used in \eqref{eq:probability that mk equals to 1}
and \eqref{eq:probability that mk equals to m} become more accurate.
Furthermore, as expected, a higher value of sensing range $R_{c}$
will result in a lower expected number of simultaneous \textit{\textcolor{black}{\emph{helper-VoI}}}\emph{
}pairs.

\begin{figure}[t]
\centering{\subfigure[Case of $R_c\geq2r_0$ ]{\label{SIMU-ANAL-Rc>=00003D2r_0}\includegraphics[width=8cm]{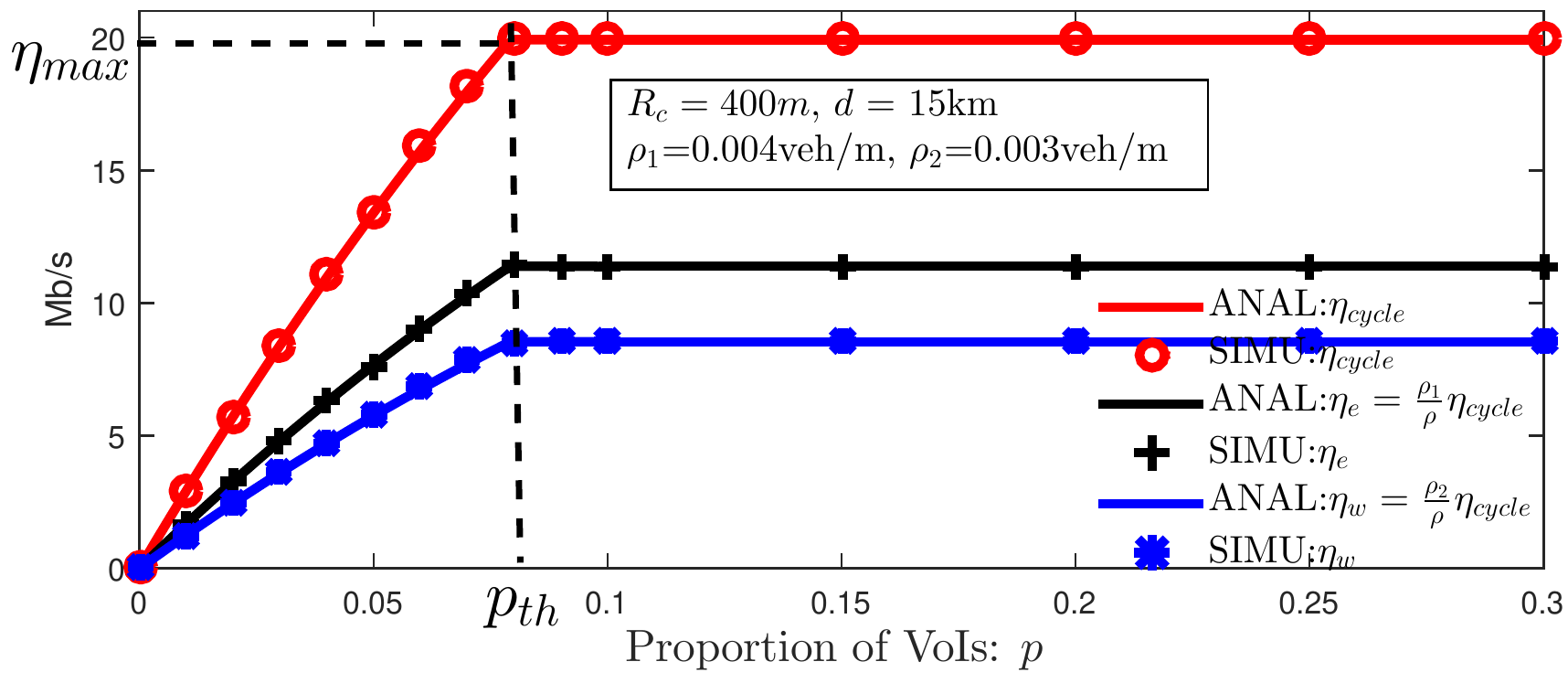}}\\
\subfigure[Case of $R_c<2r_0$]{\label{SIMU-ANAL-Rc<2r_0}\includegraphics[width=8cm]{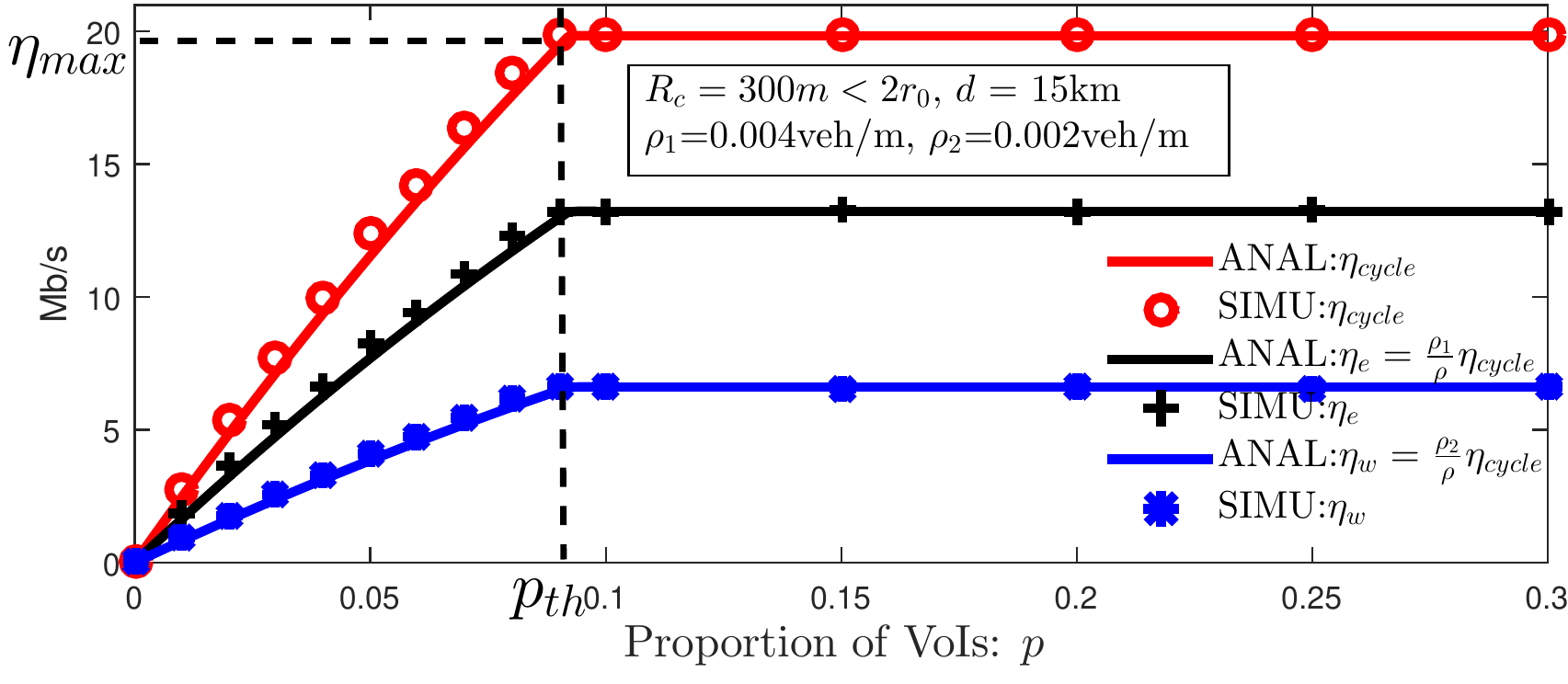}}\\
}

\caption{A comparison of the capacities achieved from one cycle by all VoIs\textit{,}
by all VoIs in the eastbound direction, and by all VoIs in the westbound
direction as a function of the proportion of VoIs $p$.}

\label{SIMU-ANAL-CAPACITY}
\end{figure}

Fig. \ref{SIMU-ANAL-CAPACITY} shows a comparison of the capacities
achieved from one cycle by all VoIs\textit{,} by all eastbound VoIs,
and by all westbound VoIs. Both Fig. \ref{SIMU-ANAL-Rc>=00003D2r_0}
and Fig. \ref{SIMU-ANAL-Rc<2r_0} show that the analytical results
match very well with simulations. Furthermore, it can be seen that
the capacity achieved by the VoIs traveling towards each direction
($\eta_{e}$ and $\eta_{w}$ respectively) is exactly proportional
to the traffic density in that direction, as predicted by Theorem
\ref{thm:The-capacity-achieved by VoIs from each direction}. 

Fig. \ref{SIMU-ANAL-CAPACITY} also reveals the relationship between
the capacity and the proportion of VoIs $p$, and shows that the capacity
increases to its maximum value when the proportion of VoIs is larger
than a certain threshold. Beyond that threshold, a further increase
in $p$ has little impact on the capacity. Specifically, as shown
in Fig. \ref{SIMU-ANAL-Rc>=00003D2r_0}, when $p$ is small, the capacity
increases sharply with an increase of $p$; however, when $p$ increases
beyond a certain threshold, e.g., $p_{th}$=0.08 in this case, a further
increase in $p$ has no impact on the capacity. This can be explained
by that when $p<p_{th}$, the number of VoIs is insufficient to retrieve
all the data received by the \textit{\textcolor{black}{\emph{helpers}}}
from their V2I communications. That is, vehicular networks offer more
data (capacity) than that can be retrieved by the VoIs and the capacity
is limited by the V2V communications between the VoIs and the \textit{\textcolor{black}{\emph{helpers}}}.
Therefore, an increase in $p$ would significantly increase the number
of simultaneous active \textit{\textcolor{black}{\emph{helper-VoI}}}\emph{
}pairs and consequently boost the capacity. However, when the proportion
of VoIs reaches a certain threshold, VoIs can retrieve almost all
the data received by the \textit{\textcolor{black}{\emph{helpers}}}
from their V2I communications. In this case, the capacity achieved
by the VoIs approaches its maximum $\eta_{max}=w_{I}(1-e^{-\rho2r_{I}})$,
which is equal to the average data rate the infrastructure point delivers
its data to all vehicles, including both VoIs and \textit{\textcolor{black}{\emph{helpers}}}. 

\begin{figure}
\centering{\includegraphics[width=8cm]{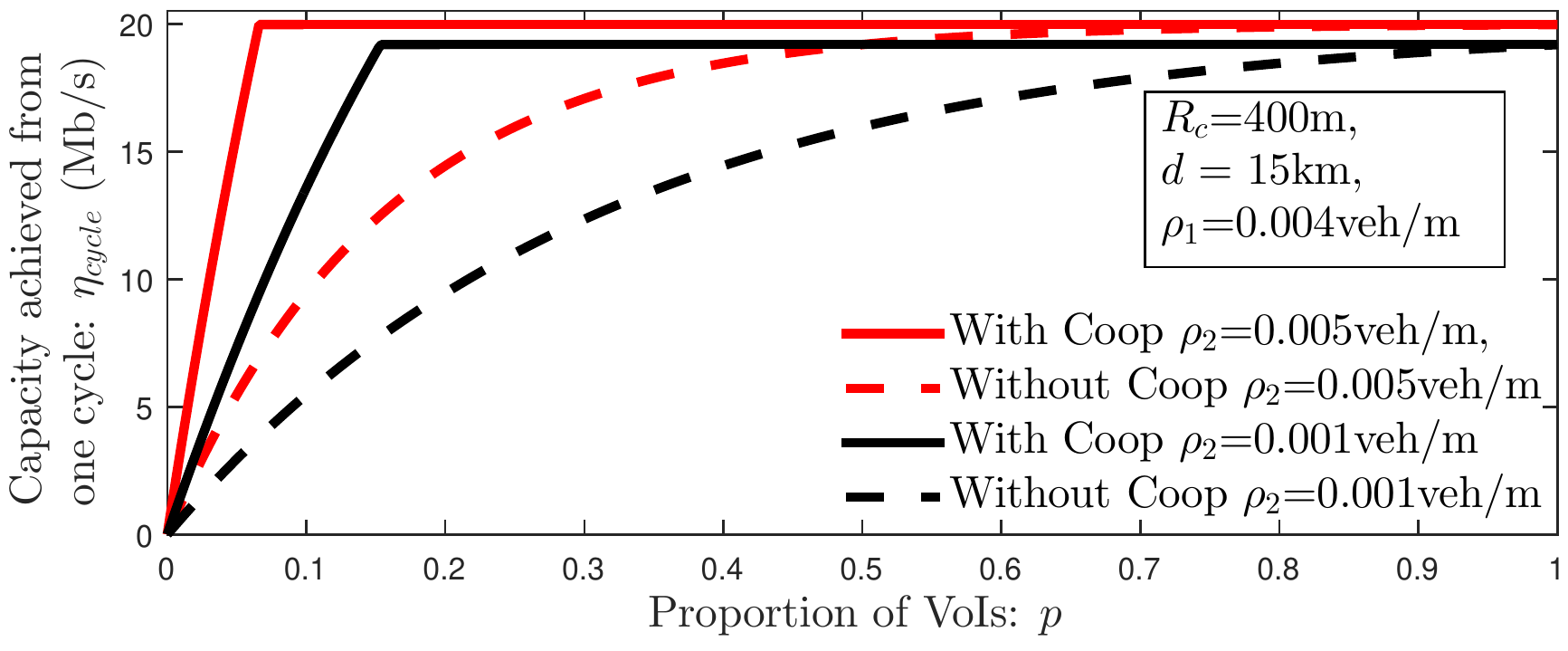}}

\caption{A comparison between the capacity achieved from one cycle as a function
of the proportion of VoIs $p$, with and without cooperative communications. }

\label{with_without_Coop}
\end{figure}

Fig. \ref{with_without_Coop} compares the capacity from one cycle
using our cooperative communication strategy (labeled as With Coop)
with its non-cooperative counterpart (labeled as Without Coop), and
shows that our cooperative communication strategy can improve the
capacity, even when there is only a small proportion of vehicles with
download requests, i.e., a small $p$. The result for the non-cooperative
counterpart is obtained by letting the VoIs only receive data from
infrastructure. It is shown that with an increase in $p$, the proposed
cooperative communication strategy becomes less effective in improving
the capacity. This is due to the fact that a larger $p$ leads to
a smaller number of \textit{\textcolor{black}{\emph{helpers}}}, which
results in a reduction in the amount of data the \textit{\textcolor{black}{\emph{helpers}}}\emph{
}can help to retrieve from the infrastructure. Thus, the contribution
to the capacity from the proposed cooperative communication strategy
becomes less significant. Furthermore, we can see that under the same
network setting, without using the cooperative communication strategy,
only when all vehicles have download requests, i.e., $p=1$, the maximum
capacity $\eta_{max}=w_{I}(1-e^{-\rho2r_{I}})$ can be achieved. In
contrast, with the cooperative communication strategy, this maximum
capacity $\eta_{max}$ can be achieved even when a small proportion
of vehicles have download requests. This validates the effectiveness
of cooperative communications to boost network performance. 

\begin{figure}
\centering{\includegraphics[width=8cm]{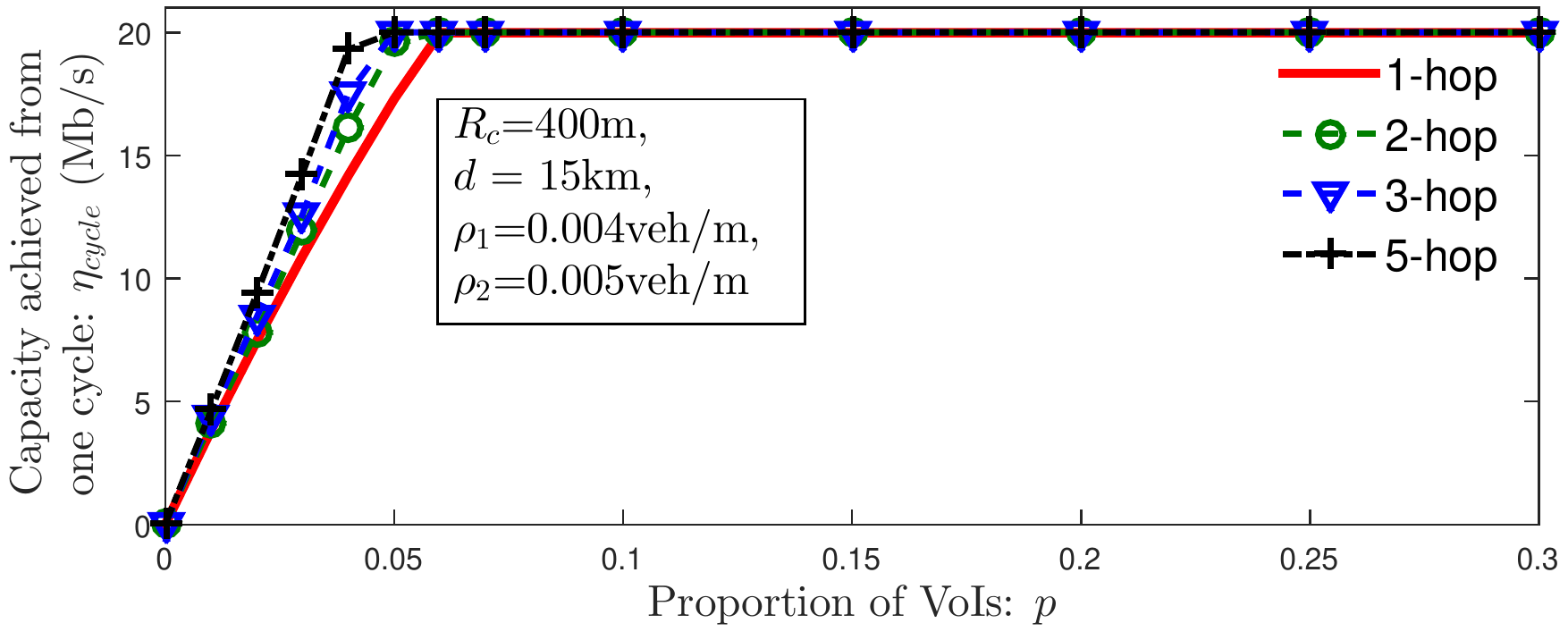}}

\caption{A comparison between capacity achieved from one cycle  when allowing
one-hop communication and multi-hop communications.}

\label{K-hop}
\end{figure}

\textcolor{black}{Fig. \ref{K-hop} compares the capacity achieved
by allowing only one-hop communication and allowing both $k$-hop
($k=2,3,5)$ V2I communications between the VoIs and infrastructure
and $k$-hop V2V communication between the VoIs and helpers. It is
shown that allowing multi-hop communications beyond one hop has little
impact on the capacity. Particularly, as pointed out in Section III-B,
in the considered scenario, allowing multi-hop V2V communications
only helps to balance the distribution of information among helpers
and does not increase the net amount of data available in the network.
The marginal increase in the achievable capacity comes from multi-hop
V2I communications between the VoIs and infrastructure, and this increase
only occurs when the proportion of VoIs, $p$, is smaller than a threshold,
e.g., $p_{th}=0.006$ in the considered scenario. }

\begin{figure}[t]
\centering{\subfigure[Achievable capacity from one \textit{cycle} under different $d$.]{\label{eta_p_d}\includegraphics[width=9.5cm]{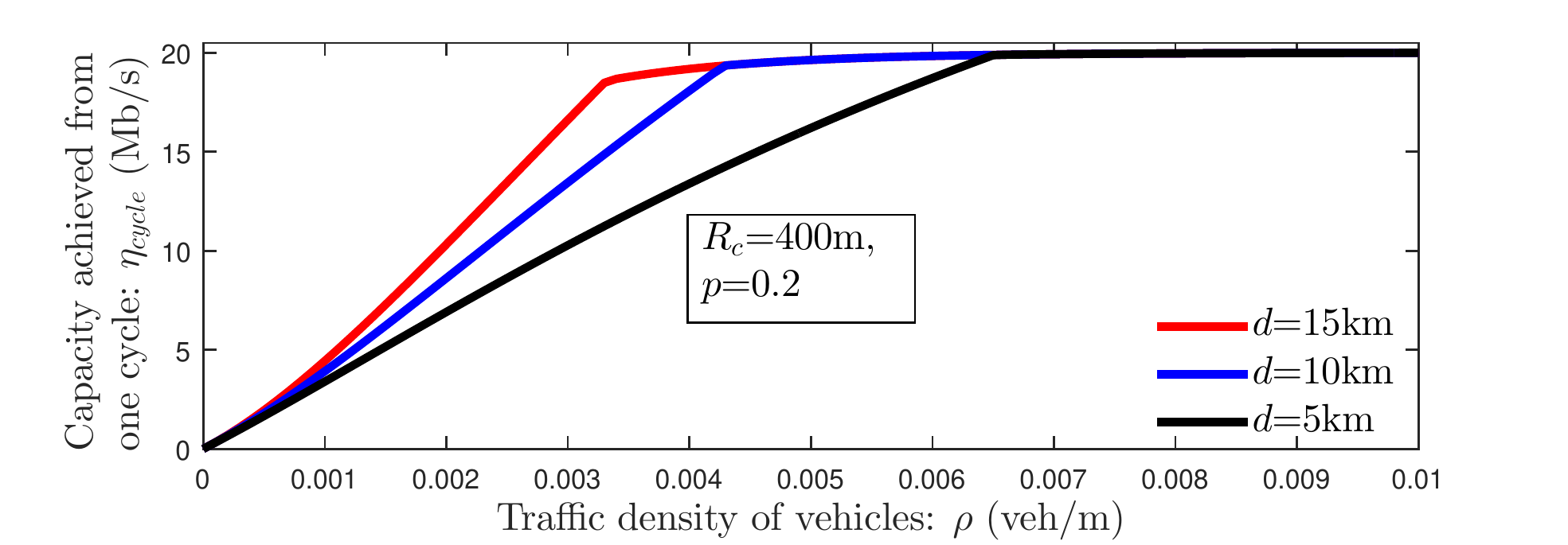}}\\
\subfigure[Total achievable capacity under different traffic density.]{\label{eta_d_rho}\includegraphics[width=9.5cm]{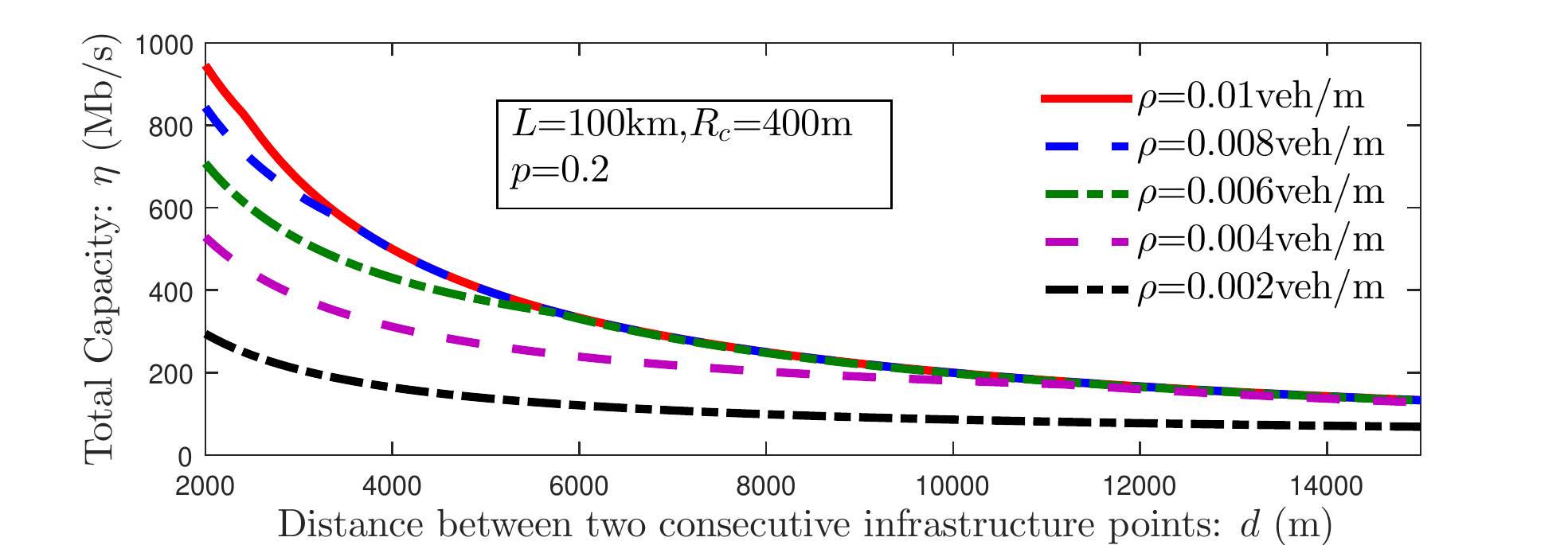}}\\
}

\caption{Relationship between capacity, distance between adjacent infrastructure
points and vehicular density.}

\label{infrastructure deploy}
\end{figure}

Fig. \ref{infrastructure deploy} demonstrates the relationship between
the capacity and the inter-infrastructure distance, and gives insight
into the optimum vehicular network infrastructure deployment in terms
of their inter-distances under different vehicular density. It is
shown in Fig. \ref{eta_p_d} that when the vehicular density $\rho$
is small, a larger $d$ will lead to a larger capacity \emph{achieved
from one cycle} because a large $d$ will increase the capacity achieved
by the VoIs from V2V communications with \textit{\textcolor{black}{\emph{helpers}}}.
However with an increase of $\rho$, the capacity achieved by the
VoIs \emph{from one cycle} under different values of $d$ differ marginally
and converge to the same maximum value. This can be explained by that
when $\rho$ is large, most of the VoIs can receive data directly
from the infrastructure, and the contribution from V2V communications
with \textit{\textcolor{black}{\emph{helpers}}} becomes less significant.
Even though an increase in $d$ would help to boost the capacity \emph{achieved
from one }\textit{cycle} when the vehicular density is small, Fig.
\ref{eta_d_rho} shows that the total achievable capacity decreases
with an increase of $d$. This is due to the fact that an increase
in $d$ on one hand brings marginal improvement on the capacity achieved
from one cycle, on the other hand, it reduces the number of \textit{\emph{cycles}},
which consequently leads to a reduction in the total capacity. Furthermore,
it can be seen that to achieve the same capacity, when the vehicular
density is larger, the inter-infrastructure distance needs to be higher.
Therefore, when determining the optimum deployment of vehicular network
infrastructure, it is important to take the vehicular density into
account, e.g., in areas where the vehicular density is usually large,
by utilizing a cooperative communication strategy, the number of infrastructure
points can be reduced. 

\section{Conclusions \label{sec:Conclusion-and-Future}}

In this paper, we analyzed the capacity of vehicular networks with
a finite traffic density adopting a cooperative communication strategy,
which utilizes V2I communications, V2V communications, mobility of
vehicles, and cooperation among vehicles and infrastructure to facilitate
the transmission. A closed-form expression of the achievable capacity
was obtained. Our result showed that the proposed cooperative strategy
can improve the capacity of vehicular networks, and the improvement
is more pronounced when the proportion of vehicles with download request
is low. Moreover, our result sheds insight into the optimum deployment
of vehicular network infrastructure and the design of cooperative
communication strategy to maximize the capacity.

\section*{Appendix A: Proof of Theorem \ref{thm:The-optimum-selection scheme}}

\begin{figure}[H]
\centering{\subfigure[Results under the proposed selection scheme $\chi_{opt}$]{\label{optimum scheme}\includegraphics[width=8cm]{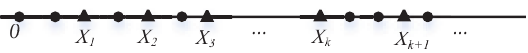}}\\
\subfigure[Results under another selection scheme $\chi^{'}$]{\label{other scheme}\includegraphics[width=8cm]{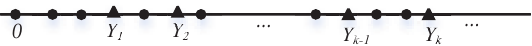}}\\
}

\caption{An illustration of the distribution of simultaneous transmitters,
where the \textcolor{black}{triangular point} represent the \textit{\textcolor{black}{\emph{helpers}}}
that are chosen as simultaneous transmitters and the \textcolor{black}{dots}
represent the \textit{\textcolor{black}{\emph{helpers}}} that are
not chosen as transmitters. }

\label{selection}
\end{figure}

Recall that we set the point to the right of infrastructure point
$I_{1}$ and at a distance $r_{I}$ to $I_{1}$ (the left boundary
point of the V2V Area) as the origin of the coordinate system, and
the right direction as the positive ($+x$) direction. Denote by $X_{k},k=1,2,...$
the location of the $i$-th transmitter (\textit{\textcolor{black}{\emph{helper}}}
of the active\textit{ }\textit{\textcolor{black}{\emph{helper-VoI}}}
pair), numbered from left to the right, under the optimum scheduling
scheme $\chi_{opt}$. Denote by $Y_{k},k=1,2,...$ the location of
the $i$-th transmitter under an arbitrary scheduling scheme $\chi^{'}$.
It follows that $X_{1}<X_{2}<\cdots<X_{k}<X_{k+1}<\cdots$ and $Y_{1}<Y_{2}<\cdots<Y_{k}<Y_{k+1}<\cdots$.
See Fig. \ref{selection} for an illustration. In the following, we
prove that $\chi_{opt}$ described in Theorem \ref{thm:The-optimum-selection scheme}
is an optimum scheduling scheme that would lead to the maximum number
of active \textit{\textcolor{black}{\emph{helper-VoI}}} pairs by recursion
that $X_{k}\leq Y_{k}$ holds for any $k=1,2,...$.

\textcolor{black}{For $k=1$, noting that according to the scheduling
scheme $\chi_{opt}$, the first transmitter is the leftmost helper
in the V2V Area that has at least one VoI within its coverage. Therefore,
it follows readily that $X_{1}\leq Y_{1}.$}

\textcolor{black}{Assuming that $X_{k}\leq Y_{k}$ when $k=n,n\geq1$,
we will show that $X_{n+1}\leq Y_{n+1}$. We consider two different
cases: $X_{n+1}\leq Y_{n}$ and $X_{n+1}>Y_{n}$:}

\textcolor{black}{(i) Case $X_{n+1}\leq Y_{n}$: in this case, it
can be readily shown that $X_{n+1}\leq Y_{n}<Y_{n+1}$. }

\textcolor{black}{(ii) Case $X_{n+1}>Y_{n}$: in this case, under
the scheduling scheme $\chi_{opt}$, the $(n+1)$-th transmitter is
the }\textcolor{black}{\emph{nearest helper}}\textcolor{black}{{} to
the right of the $n$-th transmitter satisfying simultaneous transmission
conditions: it is outside the sensing range of the $n$-th transmitter
who are located at }\textcolor{black}{\emph{$X_{n}$ and}}\textcolor{black}{{}
has at least one VoI within its transmission range that is different
from the VoI that the $n$-th transmitter transmits to. Therefore,
there is no helper}\textcolor{black}{\emph{ }}\textcolor{black}{within
road segment $(X_{n},X_{n+1})$ that can transmit simultaneously.
If $Y_{n+1}<X_{n+1}$, a contradiction must occur. Thus, $X_{n+1}\leq Y_{n+1}$. }

Therefore, $X_{k}\leq Y_{k}$ holds for any $k=1,2,...$ 

It readily follows that the number of simultaneous active \textit{\textcolor{black}{\emph{helper-VoI}}}
pairs under $\chi^{'}$ must be less than or equal to that under $\chi_{opt}$.

\section*{Appendix B: Proof of Theorem \ref{thm:Distribution of Lk}}

\textcolor{black}{Recall that we denote by $S_{k}\in[0,d-2r_{I}],k=1,2,...$
the position of the $k$-th transmitter in the V2V Area and $L_{k},k=1,2,...$
the distance between the $k$-th and the $(k+1)$-th transmitter.
}Note that each $L_{k},k=1,2,...$ are i.i.d. 

Denote by $V_{k,1}$ the first \textit{\textcolor{black}{\emph{helper}}}
located in the road segment $[S_{k}+R_{c},d-2r_{I}]$, by $V_{k,2}$
the second \textit{\textcolor{black}{\emph{helper}}}, and so on. Denote
by $l_{k,i},i=1,2,...$ the distance between \textit{\textcolor{black}{\emph{helpers}}}
$V_{k,i}$ and $V_{k,i+1}$ and by $l_{k,0}$ the distance between
\textit{\textcolor{black}{\emph{helper}}} $V_{k,1}$ and the point
$S_{k}+R_{c}$. \textcolor{black}{See Fig. \ref{illustration of the distance between two consecutive chosen helpers}
for an illustration.} As an easy consequence of the Poisson distribution
of \textit{\textcolor{black}{\emph{helpers}}}\textit{,} $l_{k,i},i=1,2,...$
follows identical and independent exponential distribution with a
mean value $\frac{1}{(1-p)\rho},$ and further due to the memoryless
property of exponential distribution \cite{Feller}, $l_{k,0}$ also
has the same distribution as $l_{k,i},i=1,2,...$. 

\begin{figure}[H]
\centering{}\includegraphics[width=8cm]{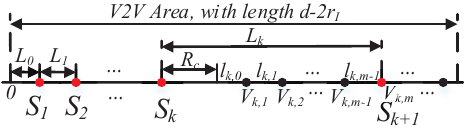} \caption{An illustration of the distribution of distances between two consecutive
simultaneous transmitters.\label{illustration of the distance between two consecutive chosen helpers-1} }
\end{figure}

\textcolor{black}{Consider the $k$-th transmitter and suppose the
$(k+1)$-th transmitter is exactly the $m_{k}$-th helper located
within the road segment $[S_{k}+R_{c},d-2r_{I}]$, where $m_{k}$
is a random integer. Note that the distribution of $m_{k}$ is independent
of $l_{k,i},i=1,2,...$, but is determined by the distribution of
VoIs within the coverage of each helper  $V_{k,1}$, $V_{k,2}$, $\cdots$
. Therefore, 
\begin{equation}
L_{k}=R_{c}+\sum_{i=0}^{m_{k}-1}l_{k,i}.\label{eq:definition of Lk}
\end{equation}
}

\textcolor{black}{From \eqref{eq:definition of Lk}, to calculate
the distribution of $L_{k}$, it remains to calculate the distribution
of $m_{k}$, $Pr(m_{k}=m)$. We compute the distribution of $m_{k}$
in the following paragraphs.}
\begin{enumerate}
\item \label{when m=00003D1}when $m=1$, \textcolor{black}{it means that
the first helper $V_{k,1}$ located within road segment $[S_{k}+R_{c},d-2r_{I}]$
is chosen as the $(k+1)$-th transmitter. This implies that there
should be at least one VoIs different from the VoI of the $k$th active
helper-VoI pair within the coverage of helper $V_{k,1}$. }

\textcolor{black}{When $R_{c}\geq2r_{0}$, the VoI of the $k$th active
helper-VoI pair cannot be possibly located within the coverage of
helper $V_{k,1}$. Therefore, when $R_{c}\geq2r_{0}$, the condition
that the helper $V_{k,1}$ is chosen as the $(k+1)$th transmitter
is that there exists at least one VoI within the coverage of helper
$V_{k,1}$, which has a length $r_{x}=2r_{0}$ (see Fig. \ref{Rc>=00003D2r_0}
for an illustration). In contrast, when $R_{c}<2r_{0}$, it may happen
that the VoI of the $k$th active helper-VoI pair is located within
the coverage of helper $V_{k,1}$ (see Fig. \ref{Rc<2r_0} for an
illustration). In this situation, helper $V_{k,1}$ may be chosen
as the $(k+1)$th transmitter iff there exists at least one VoI within
the road segment with length $r_{x}$, which starts from the position
of the VoI of $k$th active helper-VoI pair and ends at the right
boundary of the coverage of helper $V_{k,1}$, and $r_{x}<2r_{0}$. }

\begin{figure}[H]
\centering{\subfigure[ $R_c\geq2r_0$]{\label{Rc>=00003D2r_0}\includegraphics[width=8cm]{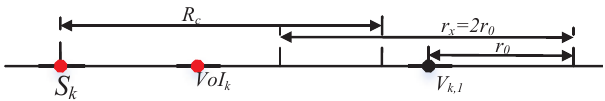}}\\
\subfigure[ $R_c<2r_0$]{\label{Rc<2r_0}\includegraphics[width=8cm]{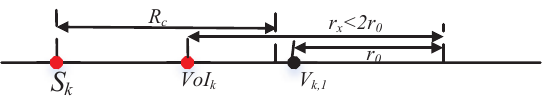}}\\
}

\caption{An illustration of the case that the \textit{\textcolor{black}{\emph{helper}}}
\textcolor{black}{$V_{k,1}$ }is chosen as the \textcolor{black}{$(k+1)$-th
transmitter}. }

\label{an illustration of r_x}
\end{figure}

\textcolor{black}{Here we approximately omit the scenario that the
VoI of $k$th active helper-VoI may be located within the coverage
of helper $V_{k,1}$. That is, we approximately consider $r_{x}$
to be equal to the length of the coverage area of helper $V_{k,1}$,
$2r_{0}$. We will use simulation later to validate the accuracy of
this approximation and its impact on our result. This approximation
allows us to write: }
\begin{align}
Pr(m_{k}=1) & =Pr\left(\exists\text{VoI within length \ensuremath{r_{x}}}\right)\nonumber \\
 & \thickapprox Pr\left(\text{\ensuremath{\exists}VoI within coverage of }V_{k,1}\right)\nonumber \\
 & =1-e^{-p\rho2r_{0}}.\label{eq:Pr(mk=00003D1)}
\end{align}

\item when $m\geq2$,\textcolor{black}{{} it means that the $m$-th helper
located within road segment $[S_{k}+R_{c},d-2r_{I}]$, $V_{k,m}$,
is chosen as the $(k+1)$-th transmitter. This implies that a) none
of the helper $V_{k,1}$, $V_{k,2}\cdots V_{k,m-1}$ satisfies the
condition to become the $(k+1)$-th transmitter, i.e., none of the
helpers $V_{k,i},i=1,...m-1$ can find a VoI that is not the same
as the VoI of the $k$-th active helper-Voi pair within their coverage;
and b) there exist at least one VoIs that is not the same as the VoI
of the $k$-th active helper-VoI pair within the coverage $V_{k,m}$.
Using the same }\textcolor{black}{\emph{approximation}}\textcolor{black}{{}
as that used previously, it can be obtained that: }
\begin{align}
 & Pr(m_{k}=m)\nonumber \\
\thickapprox & Pr\left(\text{no VoI within the coverage of }V_{k,i},i=1,...m-1\right.\nonumber \\
 & \left.\cap\text{ \ensuremath{\exists}VoI within the coverage of }V_{k,m}\right)\nonumber \\
= & Pr\left(\text{no VoI in }\left[\sum_{i=1}^{m-2}\min\left\{ l_{k,i},2r_{0}\right\} +2r_{0}\right]\right.\nonumber \\
 & \left.\cap\text{ \ensuremath{\exists}VoI in }\min\left\{ l_{k,m-1},2r_{0}\right\} \right)\nonumber \\
= & Pr\left(\text{no VoI in }\left[\sum_{i=1}^{m-2}\min\left\{ l_{k,i},2r_{0}\right\} +2r_{0}\right]\right)\times\nonumber \\
 & Pr\left(\text{\ensuremath{\exists}\ VoI in }\min\left\{ l_{k,m-1},2r_{0}\right\} \right),\label{eq:Pr(mk=00003Dm)_cal}
\end{align}
\textcolor{black}{where the last step results due to the property
that VoIs have a Poisson distribution and therefore the numbers of
VoIs in non-overlapping intervals are independent. The summation ends
at $m-2$ is due to that the total length of the coverage area of
helper $V_{k,i},i=1,...m-1$ is exactly $\sum_{i=1}^{m-2}\min\left\{ l_{k,i},2r_{0}\right\} +2r_{0}$.
}\textcolor{blue}{}\\
Define two parameters $h_{k,i}$ and $H_{k,m}$ as follows: 
\begin{equation}
h_{k,i}=\min\left\{ l_{k,i},2r_{0}\right\} ,i=1,2,...m-1\label{eq:definition of h_k,i}
\end{equation}
and 
\begin{equation}
H_{k,m}=\sum_{i=1}^{m-2}h_{k,i}+2r_{0},m=2,3,...,\label{eq:definition of H_k,m}
\end{equation}
with $H_{k,2}=2r_{0}$. Because $l_{k,i},i=1,2,...m-1$ are i.i.d.
random variables, it follows that $h_{k,i},i=1,2,...m-1$ are i.i.d.
random variables. Denote by $\bar{f}_{h_{k,i}}(x)$ and $\bar{f}_{H_{k,m}}(x)$
the pdf of random variables $h_{k,i}$ and $H_{k,m}$ respectively.
As an easy consequence of the total probability theorem, \eqref{eq:Pr(mk=00003Dm)_cal}
can be calculated by:
\begin{align}
 & Pr(m_{k}=m)\nonumber \\
\thickapprox & Pr\left(\text{no VoI in }H_{k,m}\right)\times Pr\left(\text{\ensuremath{\exists}VoI in }h_{k,m-1}\right)\nonumber \\
= & \int_{0}^{\infty}e^{-p\rho x}\bar{f}_{H_{k,m}}(x)dx\times\nonumber \\
 & \int_{0}^{\infty}\left(1-e^{-p\rho y}\right)\bar{f}_{h_{k,m-1}}(y)dy\nonumber \\
= & E\left[e^{-p\rho H_{k,m}}\right]\times\left(1-E\left[e^{-p\rho h_{k,m-1}}\right]\right).\label{eq:Pr(mk=00003Dm)_expectaion}
\end{align}
From \eqref{eq:Pr(mk=00003Dm)_expectaion}, we can see that when $m\geq2$,
the value of $Pr(m_{k}=m)$ is the product of two factors, $E\left[e^{-p\rho H_{k,m}}\right]$,
and $\left(1-E\left[e^{-p\rho h_{k,m-1}}\right]\right)$. Further
note that $E\left[e^{-p\rho H_{k,m}}\right]$ and $E\left[e^{-p\rho h_{k,m-1}}\right]$
are in the form of the moment generating functions (MGF) of the random
variables $H_{k,m}$ and $h_{k,m-1}$ respectively. For a random variable
$X$, its MGF is defined as follows \cite{Nelson95}:
\begin{equation}
M_{X}(t)\triangleq E[e^{tX}],\;\;\;\;t\in\mathbb{R}.
\end{equation}
Let $M_{H_{k,m}}(t)$ and $M_{h_{k,m-1}}(t)$ be the MFG of $H_{k,m}$
and $h_{k,m-1}$ respectively. It follows that
\begin{align}
 & Pr(m_{k}=m)\nonumber \\
\thickapprox & \left(M_{H_{k,m}}(t)\cdot\left(1-M_{h_{k,m-1}}(t)\right)\right)|_{t=-p\rho}\nonumber \\
= & \left(\prod_{i=1}^{m-2}M_{h_{k,i}}(t)\cdot M_{2r_{0}}(t)\cdot\left(1-M_{h_{k,m-1}}(t)\right)\right)|_{t=-p\rho}\nonumber \\
= & \left(\left(M_{h_{k,i}}(t)\right)^{m-2}\cdot e^{2r_{0}t}\cdot\left(1-M_{h_{k,i}}(t)\right)\right)|_{t=-p\rho}\label{eq:Pr(mk=00003Dm) as a function of MGF}
\end{align}
\textcolor{black}{where the second step results because of the fact
that the random variable $H_{k,m}$ is the sum of $m-2$ i.i.d. random
variables $h_{k,i}$ and a constan $2r_{0}$. }\\
It remains to calculate the MGF of $h_{k,i}$. Noting that $l_{k,i},i=1,2,...m-1$
has an exponential distribution: $f_{l_{k,i}}(x)=(1-p)\rho e^{-(1-p)\rho x}$.
The pdf of $h_{k,i}=\min\left\{ l_{k,i},2r_{0}\right\} $, $\bar{f}_{h_{k,i}}(x)$,
can be calculated as follows:
\begin{equation}
\bar{f}_{h_{k,i}}(x)=\begin{cases}
(1-p)\rho e^{-(1-p)\rho x} & x<2r_{0}\\
e^{-(1-p)\rho2r_{0}}\delta(x-2r_{0}) & x\geq2r_{0}
\end{cases}\label{eq:mixed p.d.f of h_k,i}
\end{equation}
where $\delta(x)$ is a delta function:
\begin{align}
\delta(x)= & \begin{cases}
1, & x=0\\
0, & x\neq0
\end{cases}\label{eq:definition of delta function}
\end{align}
Using \eqref{eq:mixed p.d.f of h_k,i}, the MGF of $h_{k,i}$ can
be obtained as follows:
\begin{align}
M_{h_{k,i}}(t)= & E[e^{t\cdot h_{k,i}}]\nonumber \\
= & \int_{0}^{\infty}e^{tx}\bar{f}_{h_{k,i}}(x)dx\nonumber \\
= & \int_{0}^{2r_{0}}e^{tx}(1-p)\rho e^{-(1-p)\rho x}dx+e^{t2r_{0}}e^{-(1-p)\rho2r_{0}}\nonumber \\
= & \frac{te^{\left(t-(1-p)\rho\right)2r_{0}}-(1-p)\rho}{t-(1-p)\rho}.\label{eq:mgf of h_k,i}
\end{align}
Combing \eqref{eq:Pr(mk=00003Dm) as a function of MGF} and \eqref{eq:mgf of h_k,i}
and simplifying it, for $m\geq2$, we have 
\begin{equation}
Pr(m_{k}=m)\thickapprox e^{-p\rho2r_{0}}\cdot c_{1}^{m-2}\left(1-c_{1}\right)\label{eq:Pr(mk=00003Dm)}
\end{equation}
where $c_{1}=M_{h_{k,i}}(t)|_{t=-p\rho}=1-p+pe^{-\rho2r_{0}}.$
\end{enumerate}
Combing the above result, we have the cdf of $L_{k},k\geq1$, denoted
by $F_{L_{k}}(x)$, as follows:
\begin{align}
 & F_{L_{k}}(x)\nonumber \\
= & Pr(L_{k}\leq x)\nonumber \\
= & \sum_{m=1}^{\infty}Pr(R_{c}+\sum_{i=0}^{m_{k}-1}l_{k,i}\leq x|m_{k}=m)Pr(m_{k}=m)\nonumber \\
= & \sum_{m=1}^{\infty}Pr(\sum_{i=0}^{m-1}l_{k,i}\leq x-R_{c})Pr(m_{k}=m)\nonumber \\
= & \sum_{m=1}^{\infty}\sum_{n=m}^{\infty}\frac{e^{-(1-p)\rho(x-R_{c})}\left[(1-p)\rho(x-R_{c})\right]^{n}}{n!}Pr(m_{k}=m)\label{eq:CDF of Lk}
\end{align}
\textcolor{black}{where the second step is obtained by putting \eqref{eq:definition of Lk}
into $L_{k}$ and using the total probability theorem and the last
step is obtained by that the distribution of $\sum_{i=0}^{m-1}l_{k,i}$
is independent of the distribution of $m_{k}$.}

From \eqref{eq:CDF of Lk} and with $f_{L_{k}}(x)=\frac{dF_{L_{k}}(x)}{dx}$,
we have the \textit{\emph{pdf }}of $L_{k}$ shown as \eqref{eq:pdf of Lk},
which completes the proof.

\section*{Appendix C: Proof of Theorem \ref{thm:The-capacity-achieved by VoIs from each direction} }

To calculate the capacity achieved by the eastbound and westbound
VoIs, we will analyze the V2V communications and V2I communications
in one cycle area separately. 

1). V2V communications:

\textcolor{black}{Recall that under our optimum helper-VoI scheduling
algorithm $\chi_{opt}$ for the V2V communications proposed in Theorem
\ref{thm:The-optimum-selection scheme}, for any two randomly chosen
helper-VoI pairs at a randomly chosen time slot, the travel direction
of the VoI in one pair is independent of the travel direction of the
helper in the same pair, and is also independent of the travel direction
of the VoI in the other pair. Therefore, at any randomly chosen time
slot, the proportion of the helper-VoI pairs whose VoI is eastbound
(westbound) is equal to the probability that the VoI of a randomly
chosen helper-VoI pair is eastbound (westbound), denoted by $P_{Ve}$
($P_{Vw}$). Obviously, $P_{Ve}+P_{Vw}=1$. It follows that the expected
amount of data the eastbound and westbound VoIs can receive through
V2V communications during time period $\left[0,t\right]$ from one
cycle area, denote by $D_{V2Ve}(t)$ and $D_{V2Vw}(t)$ respectively,
can be calculated by $D_{V2Ve}(t)=P_{Ve}\cdot D_{V2V}(t)$ and $D_{V2Vw}(t)=P_{Vw}\cdot D_{V2V}(t)$,
where $D_{V2V}(t)$ is the expected amount of data received by all
the VoIs given in  \eqref{eq:definition of capcity achievable from one cycle}.
Therefo}re, the capacity achieved by the eastbound and westbound VoIs
through V2V communications from one cycle area are respectively: 
\begin{equation}
\lim_{t\rightarrow\infty}\frac{D_{V2Ve}(t)}{t}=P_{Ve}\lim_{t\rightarrow\infty}\frac{D_{V2V}(t)}{t},\label{eq:capacity achieved by VoIs from eastbound}
\end{equation}
 and 
\begin{equation}
\lim_{t\rightarrow\infty}\frac{D_{V2Vw}(t)}{t}=P_{Vw}\lim_{t\rightarrow\infty}\frac{D_{V2V}(t)}{t}.\label{eq:capacity achieved by VoIs from westbound}
\end{equation}

\begin{figure}
\begin{centering}
\includegraphics[width=8.5cm]{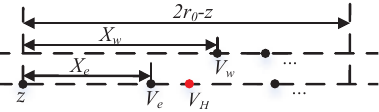}
\par\end{centering}
\caption{An illustration of the coordinate system, the location of the randomly
chosen transmitter, and the left-most VoI from each direction that
are located at the right of origin. }

\label{VoI-selection}
\end{figure}

\textcolor{black}{In the following, we will calculate $P_{Ve}$ and
$P_{V_{w}}$. Suppose }\textit{\textcolor{black}{\emph{helper}}}\textcolor{black}{{}
$V_{H}$ is one of the simultaneous transmitters at a randomly chosen
time slot. Noting that for a randomly chosen helper-VoI pair, the
travel direction of its VoI is irrelevant to the travel direction
of its helper, but only dependent on the original distribution of
VoIs in each direction. Therefore, without loss of generality, we
assume }\textit{\textcolor{black}{\emph{that helper}}}\textcolor{black}{{}
$V_{H}$ travels eastbound.  Recall that we designate the east (right)
direction as $+x$ direction. Here, designate the point to the $-x$
direction of }\textit{\textcolor{black}{\emph{helper}}}\textcolor{black}{{}
$V_{H}$ and at a distance $r_{0}$ to $V_{H}$ as the origin of the
coordinate system. Denote by $z$ the location of the point from which
the }\textit{\textcolor{black}{\emph{helper}}}\textcolor{black}{{} $V_{H}$
starts to choose its receiver (VoI), i.e., the }\textit{\textcolor{black}{\emph{helper}}}\textcolor{black}{{}
$V_{H}$ chooses its receiver within road segment $[z,2r_{0}]$. Therefore,
according to the scheduling scheme $\chi_{opt}$, $z$ is equal to
$0$ if the VoI of the previous }\textit{\textcolor{black}{\emph{helper-VoI}}}\textcolor{black}{{}
pair is not located within the coverage of }\textit{\textcolor{black}{\emph{helper}}}\textcolor{black}{{}
$V_{H}$, otherwise $z>0$. }See Fig. \ref{VoI-selection} for an
illustration. 

Denote the left-most eastbound VoI located at the right side of $z$
by $V_{e}$, and its location by $X_{e}$. Further denote the left-most
westbound VoI located at the right of $z$ by $V_{w}$, and its location
by $X_{w}$. Noting that eastbound and westbound VoIs follow Poisson
distributions with densities $p\rho_{1}$ and $p\rho_{2}$ respectively,
it follows that $X_{e}$ and $X_{w}$ are exponentially distributed:
$f_{X_{e}}(x)=p\rho_{1}e^{-p\rho_{1}x}$ and $f_{X_{w}}(x)=p\rho_{2}e^{-p\rho_{2}x}$.
\textcolor{black}{Given that there is at least one VoI within road
segment $[z,2r_{0}]$ (otherwise}\textit{\textcolor{black}{{} }}\textit{\textcolor{black}{\emph{helper}}}\textcolor{black}{{}
$V_{H}$ can not be one of the simultaneous transmitters), VoI $V_{e}$
can be chosen as the receiver of }\textit{\textcolor{black}{\emph{helper}}}\textcolor{black}{{}
$V_{H}$ iff when $X_{e}\leq X_{w}$. Therefore, we have the probability
that the receiver of the }\textit{\textcolor{black}{\emph{helper}}}\textcolor{black}{{}
$V_{H}$ travels towards east as follows (conditioned on $z$ is fixed):}
\begin{align}
 & P_{Ve}\nonumber \\
= & Pr(X_{e}\leq X_{w}|\text{there exists VoI in }[z,2r_{0}])\nonumber \\
= & \frac{Pr(X_{e}\leq X_{w},\text{there exists VoI in }[z,2r_{0}])}{Pr(\text{there exists VoI in }[z,2r_{0}])}\nonumber \\
= & \frac{\int_{0}^{2r_{0}-z}Pr(X_{e}\leq x)f_{X_{w}}(x)dx}{1-e^{-p\rho(2r_{0}-z)}}+\nonumber \\
 & \frac{\int_{2r_{0}-z}^{\infty}Pr(X_{e}\leq2r_{0}-z)f_{X_{w}}(x)dx}{1-e^{-p\rho(2r_{0}-z)}}\nonumber \\
= & \frac{\rho_{1}}{\rho_{1}+\rho_{2}},\label{eq:Probability that the VoI in one pair comes from eastbound}
\end{align}
\textcolor{black}{where the second step results by using Bayes' theorem;
the third step is obtained by using total probability theorem. In
addition, the result of \eqref{eq:Probability that the VoI in one pair comes from eastbound}
is irrelevant to the random variable $z$. Straightforwardly, we have:
}
\begin{equation}
P_{Vw}=1-P_{Ve}=\frac{\rho_{2}}{\rho_{1}+\rho_{2}}.\label{eq:Probability that the VoI in one pair comes from westbound}
\end{equation}

Plugging \eqref{eq:Probability that the VoI in one pair comes from eastbound}
and \eqref{eq:Probability that the VoI in one pair comes from westbound}
into \eqref{eq:capacity achieved by VoIs from eastbound} and \eqref{eq:capacity achieved by VoIs from westbound}
respectively, we can conclude that the expected amount of data received
by the eastbound and westbound VoIs through V2V communications are
proportional to their respective traffic densities. 

2). V2I communications: 

\textcolor{black}{Denote by $P_{Ie}$ and $P_{Iw}$ respectively the
probability that at a randomly chosen time slot, the receiver of a
VoIs' V2I communication travels towards east and west. Using the same
method above for analyzing the V2V communications (the infrastructure
point in this case corresponds to the randomly chosen transmitter
$V_{H}$, and coverage area of infrastructure $2r_{I}$ corresponds
the VoI choosing area $2r_{0}-z$), it is ready to have $P_{Ie}=\frac{\rho_{1}}{\rho_{1}+\rho_{2}}$,
and $P_{Iw}=\frac{\rho_{2}}{\rho_{1}+\rho_{2}}$. Therefore, the capacity
achieved by the eastbound and westbound VoIs respectively through
V2I communications in one cycle area, can be obtained by}
\begin{equation}
\lim_{t\rightarrow\infty}\frac{D_{V2Ie}(t)}{t}=\frac{\rho_{1}}{\rho_{1}+\rho_{2}}\cdot\lim_{t\rightarrow\infty}\frac{D_{V2I}(t)}{t},\label{eq:capacity achieved by VoIs from eastbound from V2I communications}
\end{equation}
and
\begin{equation}
\lim_{t\rightarrow\infty}\frac{D_{V2Iw}(t)}{t}=\frac{\rho_{2}}{\rho_{1}+\rho_{2}}\cdot\lim_{t\rightarrow\infty}\frac{D_{V2I}(t)}{t}.\label{eq:capacity achieved by VoIs from westbound from V2i communications}
\end{equation}
\textcolor{black}{where $D_{V2Ie}(t)$ and $D_{V2Iw}(t)$ are respectively
the expected amount of data the eastbound and westbound VoIs can receive
through V2I communications during time period $t$ from one cycle}\textit{\textcolor{black}{{}
}}\textcolor{black}{area}\textit{\textcolor{black}{.}}

\textcolor{black}{Noting that both the analysis in 1) and 2) show
that the amount of data received from V2V communications and V2I communications
by the eastbound and westbound VoIs are proportional to their respective
traffic densities. Therefore, the capacity achieved from one cycle
area by the eastbound and westbound VoIs, are also proportional to
their traffic densities respectively, which finalizes the proof.}

\begin{IEEEbiography}{Jieqiong Chen}
(S'16) received the Bachelor\textquoteright s degree in Engineering
from Zhejiang University, Zhejiang, China, in 2012, and she is currently
working toward the Ph.D. degree in engineering at the University of
Technology Sydney, Sydney, Australia. Her research interests in the
area of vehicular networks.
\end{IEEEbiography}

\begin{IEEEbiography}{Guoqiang Mao}
(S'98-M'02-SM'07) received PhD in telecommunications engineering
in 2002 from Edith Cowan University. Between 2002 and 2013, he was
an Associate Professor at the School of Electrical and Information
Engineering, the University of Sydney. He currently holds the position
of Professor of Wireless Networking, Director of Center for Real-time
Information Networks at the University of Technology, Sydney. He has
published more than 100 papers in international conferences and journals,
which have been cited more than 3000 times. His research interest
includes intelligent transport systems, applied graph theory and its
applications in networking, wireless multi-hop networks, wireless
localization techniques and network performance analysis. He is an
Editor of IEEE Transactions on Vehicular Technology and IEEE Transactions
on Wireless Communications, and a co-chair of IEEE Intelligent Transport
Systems Society Technical Committee on Communication Networks. 
\end{IEEEbiography}

\begin{IEEEbiography}{Changle Li}
(M'09) received the B.E. degree in microwave telecommunication engineering
and the M.E. and Ph.D. degrees in communication and information system
from Xidian University, Xi?an, China, in 1998, 2001, and 2005, respectively.
From 2006 to 2007, he was with the Department of Computer Science,
University of Moncton, Moncton, NB, Canada, as a Postdoctoral Researcher.
From 2007 to 2009, he was an expert researcher at the National Institute
of Information and Communications Technology, Japan. He is currently
a Professor with the State Key Laboratory of Integrated Services Networks,
Xidian University. His research interests include intelligent transportation
systems, vehicular networks, mobile ad hoc networks, and wireless
sensor networks. 
\end{IEEEbiography}

\begin{IEEEbiography}{Weifa Liang}
(M\textquoteright 99\textendash SM\textquoteright 01) received the
B.Sc. degree from Wuhan University, China, in 1984, the M.E. degree
from The University of Science and Technology of China in 1989, and
the Ph.D. degree from The Australian National University in 1998,
all in computer science. He is currently a Professor with the Research
School of Computer Science, The Australian National University. His
research interests include design and analysis of energyefficient
routing protocols for wireless ad hoc and sensor networks, cloud computing,
software-defined networking, design and analysis of parallel and distributed
algorithms, approximation algorithms, combinatorial optimization,
and graph theory. 
\end{IEEEbiography}

\begin{IEEEbiography}{Degan Zhang}
(M\textquoteright 01) was born in 1969. He received the Ph.D. degree
from Northeastern University, Shenyang, China. He is currently a Professor
with Tianjin Key Lab of Intelligent Computing and Novel Software Technology,
Key Lab of Computer Vision and System, Ministry of Education, Tianjin
University of Technology, Tianjin 300384, China. His research interests
include wireless sensor networks and industrial applications.
\end{IEEEbiography}


\begin{thebibliography}{10}
\bibitem{Zheng15}K. Zheng, et al., ``Heterogeneous Vehicular Networking:
A Survey on Architecture, Challenges, and Solutions,'' \textit{IEEE
Commun. Surveys Tuts.}, vol. 17, no. 4, pp. 2377-2396, Fourth Quarter,
2015. 

\bibitem{Ilarri15} S. Ilarri, T. Delot, R. Trillo-Lado, ``A Data
Management Perspective on Vehicular Networks,'' \textit{IEEE Commun.
Surveys Tuts.}, vol. 17, no. 4, pp. 2420-2460, Fourth Quarter, 2015. 

\bibitem{Li2014} Y. Li, X. Zhu, D. Jin and D. Wu, ``Multiple Content
Dissemination in Roadside-Unit-Aided Vehicular Opportunistic Networks,''
\textit{IEEE Trans. Veh. Technol.}, vol. 63, no. 8, pp. 3947-3956,
Oct. 2014.

\bibitem{Ning14} N. Lu, N. Cheng, N. Zhang, X. Shen and J. W. Mark,
``Connected Vehicles: Solutions and Challenges,'' \emph{IEEE Internet
Things J.}, vol 1, no. 4, pp. 289-299, Aug. 2014.

\bibitem{Kenny11}J. B. Kenney, ``Dedicated Short-Range Communications
(DSRC) Standards in the United States,'' \textit{Proceedings of the
IEEE}, vol. 99, no. 7, pp. 1162-1182, 2011. 

\bibitem{Morgan10}Y. L. Morgan, ``Notes on DSRC \& WAVE Standards
Suite: Its Architecture, Design, and Characteristics,'' \textit{IEEE
Commun. Surveys Tuts.}, vol.12, no, 4, pp. 504-518, Fourth Quarter,
2010. 

\bibitem{Khabazian13} M. Khabazian, S. Aissa, and M. Mehmet-Ali,
``Performance Modeling of Safety Messages Broadcast in Vehicular
Ad Hoc Networks,'' \emph{IEEE Trans. Intell. Transp. Syst.}, vol.
14, no. 1, pp. 380-387, Mar, 2013.

\bibitem{Zhang2014} Z. Zhang, G. Mao, and B. D. O. Anderson, ``Stochastic
Characterization of Information Propagation Process in Vehicular Ad
hoc Networks,'' \emph{IEEE Trans. Intell. Transp. Syst.}, vol. 15,
no. 1, pp. 122-135, Feb. 2014.

\bibitem{Du15}L. Du, and H. Dao, ``Information Dissemination Delay
in Vehicle-to-Vehicle Communication Networks in a Traffic Stream'',
\emph{IEEE Trans. Intell. Transp. Syst.}, vol 16, no. 1, pp. 66-80,
Feb. 2015.

\bibitem{Shahidi} R. Shahidi, and M. H. Ahmed, ``Probability Distribution
of End-to-End Delay in a Highway VANET,'' \textit{IEEE Commun. Lett.},
vol. 18, no. 3, pp. 443-446, Mar. 2014. 

\bibitem{Gupta}P. Gupta and P. Kumar, ``The capacity of wireless
networks,'' \emph{IEEE Trans. Inf. Theory}, vol. 46, no. 2, pp. 388-404,
2000.

\bibitem{Gross02} M. Grossglauser and D. Tse, ``Mobility increase
the capacity of ad hoc wireless networks,'' \emph{IEEE/ACM Trans.
Netw.}, vol. 10, no. 4, pp. 477-486, Aug. 2002. 

\bibitem{Mao13} G. Mao, Z. Lin, X. Ge, and Y. Yang, ``Towards a
Simple Relationship to Estimate the Capacity of Static and Mobile
Wireless Networks,'' \emph{IEEE Trans. Wireless Commun.,} vol. 12,
no. 9, pp. 3883-3895, Aug. 2013.

\bibitem{Dai2016}H. N. Dai, R. C. W. Wong, and H. Wang, ``On Capacity
and Delay of Multi-channel Wireless Networks with Infrastructure Support,''
\textit{IEEE Trans. Veh. Technol.}, 2016, to appear. 

\bibitem{Wang14} M. Wang, et al., ``Asymptotic Throughput Capacity
Analysis of VANETs Exploiting Mobility Diversity,'' \textit{IEEE
Trans. Veh. Technol.}, vol. 64, no. 9, pp. 4187 - 4202, Sep. 2015.

\bibitem{Huang2015}Y. Huang, M. Chen, Z. Cai, X. Guan, T. Ohtsuki,
and Y. Zhang, ``Graph Theory Based Capacity Analysis for Vehicular
Ad Hoc Networks,'' in \textit{IEEE GLOBECOM,} 2015. 

\bibitem{CHEN16}J. Chen, A. Zafar, G. Mao, C. Li, ``On the Achievable
Throughput of Cooperative Vehicular Networks,'' in \textit{Proceedings
of IEEE ICC 2016}. 

\bibitem{Zhu15} W. Zhu, D. Li, and W. Saad, ``Multiple Vehicles
Collaborative Data Download Protocol via Network Coding,'' \textit{IEEE
Trans. Veh. Technol.}, vol. 64, no. 4, pp. 1607-1619, Apr. 2015.

\bibitem{Haibo14} H. Zhou et.al, ``ChainCluster: Engineering a Cooperative
Content Distribution Framework for Highway Vehicular Communications,''
\emph{IEEE Trans. Intell. Transp. Syst.}, vol. 15, no. 6, pp. 2644-2657,
Dec. 2014.

\bibitem{Das2016} B. Das, S. Misra, U. Roy, ``Coalition Formation
for Cooperative Service-Based Message Sharing in Vehicular Ad Hoc
Networks,'' \textit{IEEE Trans. Parallel Distrib. Syst.}, vol. 27,
no. 1, pp. 144-156, Jan. 2016. 

\bibitem{Liu2016}K. Liu, et al., ``Cooperative Data Scheduling in
Hybrid Vehicular Ad Hoc Networks: VANET as a Software Defined Network,''
\emph{IEEE/ACM Trans. Netw.}, vol. 24, no. 3, pp. 1759 - 1773, Jun.
2016.

\bibitem{Zhang2013}D. Zhang and C. Yeo,\textquotedblleft Enabling
efficient WiFi-based vehicular content distribution,\textquotedblright{}
\textit{IEEE Trans. Parallel Distrib. Syst.}, vol. 24, no. 3, pp.
479-492, Mar. 2013.

\bibitem{Kim15} R. Kim, H. Lim, and B. Krishnamachari, ``Prefetching-Based
Data Dissemination in Vehicular Cloud Systems,'' \textit{IEEE Trans.
Veh. Technol.}, vol. 65, no. 1, pp. 292 - 306, Jan. 2016.

\bibitem{Mauri16}G. Mauri, M. Gerla, F. Bruno, M. Cesana, and G.
Verticale, ``Optimal Content Prefetching in NDN Vehicle-to-Infrastructure
Scenario,'' \textit{IEEE Trans. Veh. Technol.}, 2016, to appear. 

\bibitem{Mershad12} K. Mershad, H. Artail, and M. Gerla, ``We Can
Deliver Messages to Far Vehicles,'' \emph{IEEE Trans. Intell. Transp.
Syst.}, vol. 13, no. 3, pp. 1099-1115, Sep. 2012.

\bibitem{Si16}P. Si, Y. He, H. Yao, R. Yang, and Y. Zhang, ``DaVe:
Offloading Delay-Tolerant Data Traffic to Connected Vehicle Networks,''
\textit{IEEE Trans. Veh. Technol.}, vol. 65, no. 6, pp. 3941 - 3953,
Jun. 2016. 

\bibitem{Wang16} Y. Wang, Y. Liu, J. Zhang, H. Ye, and Z. Tan, ``Cooperative
Store-Carry-Forward Scheme for Intermittently Connected Vehicular
Networks,'' \textit{IEEE Trans. Veh. Technol.},'' 2016, to appear. 

\bibitem{Abboud14} K. Abboud and W. Zhuang, ``stochastic Analysis
of a Single-Hop Communication Link in Vehicular Ad Hoc Networks,''
\emph{IEEE Trans. Intell. Transp. Syst.}, vol. 15, no. 5, pp. 2297-2307,
Oct. 2014.

\bibitem{Zhang12} W. Zhang, et al., ``Multi-Hop Connectivity Probability
in Infrastructure-Based Vehicular Networks,'' \emph{IEEE J. Sel.
Areas Commun.}, vol. 30, no. 4, pp. 740-747, May. 2012.

\bibitem{Wis2007} N. Wisitpongphan, B. Fan, P. Mudalige, V. Sadekar,
and O. Tonguz, ``Routing in Sparse Vehicular Ad Hoc Wireless Networks,''
\emph{IEEE J. Sel. Areas Commun.}, vol. 25, no. 8, pp. 1538-1556,
Oct. 2007.

\bibitem{Reis14} A. B. Reis, et al., ``Deploying Roadside Units
in Sparse Vehicular Networks: What Really Works and What Does Not,''
\textit{IEEE Trans. Veh. Technol.}, vol. 63, no. 6, pp. 2794-2806,
Jul. 2014.

\bibitem{Ruixue13}R. Mao and G. Mao, ``Road traffic density estimation
in vehicular networks,'' in \textit{IEEE Wireless Communications
and Networking Conference (WCNC)}, 2013. 

\bibitem{Nelson95}R. Nelson, \textit{Probability, Stochastic Processes,
and Queueing Theory: The Mathematics of Computer Performance Modeling,
}New York: Springer- Verlag, 1995.

\bibitem{Ge15}X. Ge, S. Tu, T. Han, Q. Li and G. Mao, \textquotedblleft Energy
Efficiency of Small Cell Backhaul Networks Based on Gauss-Markov Mobile
Models,\textquotedblright{} \textit{\textcolor{black}{IET Networks}},
Vol. 4, No. 2, pp. 158-167, 2015.

\bibitem{Mao2006}G. Mao, B. D. O. Anderson, and B. Fidan, ``Online
Calibration of Path Loss Exponent in Wireless Sensor Networks,''
in \textit{IEEE Globecom}, 2006. 

\bibitem{Mao09}G. Mao, and B. D. O. Anderson, ``Graph Theoretic
Models and Tools for the Analysis of Dynamic Wireless Multihop Networks,''
in \textit{IEEE Wireless Communications and Networking Conference
(WCNC)}, 2009. 

\bibitem{Cheng15}J. Cheng, J. Cheng, M. Zhou, F. Liu, S. Gao, and
C. Liu, ``Routing in Internet of Vehicles: A Review,'' \textit{IEEE
Trans. Intell. Transp. Syst.}, vol. 16, no. 5, pp. 2339-2352, Oct.
2015.

\bibitem{Kumar04} A. Ashish and P. R. Kumar, ``Capacity bounds for
ad hoc and hybrid wireless networks,'' \textit{ACM SIGCOMM Computer
Communications Review}, vol. 34, no. 3, pp. 71-81, Jul. 2004.

\bibitem{Yang15}H. Yang, and W. Jin, ``Instantaneous communication
throughputs of vehicular ad hoc networks,'' \textit{Transportation
Research Part C: Emerging Technologies,} vol. 53, pp. 19-34, 2015. 

\bibitem{WangPeng15} P. Wang, G. Mao, Z. Li, X. Ge and B. D. O. Anderson,
``Network Coding based Wireless Broadcast with Performance Guarantee,''
\emph{IEEE Trans. Wireless Commun.}, Vol. 14, No. 1, pp. 532 - 544,
Jan. 2015.

\bibitem{Gallager13} R. G. Gallager, \textit{Stochastic Processes:
Theory for Applications}. Cambridge University Press, 2013.

\bibitem{Baccelli09}F. Baccelli and B. B\l aszczyszyn, \textit{Stochastic
Geometry and Wireless Networks Volume II Applications}, 2009.

\bibitem{Feller}W. Feller, \textit{An Introduction to Probability
Theory and Its Applications, Volume II}, 1971.
\end{thebibliography}
\end{document}